\documentclass[preprint]{sig-alternate-05-2015}
\toappear{To appear in ACM Asia Conference on Computer and Communications Security (ASIACCS) 2017.}
\usepackage[utf8]{inputenc}
\usepackage{amsmath}
\usepackage{amsfonts}
\usepackage{amssymb}
\usepackage{graphicx}
\graphicspath{{./}}
\usepackage{booktabs}
\usepackage{tabu}
\usepackage{changepage}
\usepackage{pifont}
\usepackage{float}
\usepackage{flafter}
\usepackage[keeplastbox]{flushend}
\usepackage{xspace}
\usepackage{paralist}

\usepackage[backend=bibtex]{biblatex}
\bibstyle{abbrv}
\AtEveryBibitem{\clearlist{issn}}
\AtEveryBibitem{\clearfield{volume}}
\AtEveryBibitem{\clearfield{isbn}}
\AtEveryBibitem{\clearfield{doi}}

\newcommand{\attack}{{\em S\&T attack}\xspace}
\newcommand{\att}{{\em S\&T}\xspace}
\newcommand{\attackfull}{{\em Skype \& Type attack}\xspace}
\newcommand{\target}{{\sf\small target-device}\xspace}
\newcommand{\Target}{{\sf\small Target-device}\xspace}

\newcommand{\Targetbf}{{\sf\bf Target-device}\xspace}
\newcommand{\targetfootnote}{{\sf\footnotesize target-device}\xspace}
\newcommand{\attdev}{{\sf\small attack-device}}
\newcommand{\tartxt}{{\sf\small target-text}}

\newcommand{\descr}[1]{\medskip\noindent\textbf{#1}}
\renewcommand{\paragraph}{\descr}

\begin{document}
	
\clubpenalty=10000
\widowpenalty = 10000

\title{Don't Skype \& Type! \\ Acoustic Eavesdropping in Voice-Over-IP}

\numberofauthors{4}

\author{
\alignauthor
Alberto Compagno\\
\affaddr{Sapienza University of Rome}\\
\email{compagno@di.uniroma1.it}
\alignauthor
Mauro Conti\\
\affaddr{University of Padua}\\
\email{conti@math.unipd.it}
\and
\alignauthor Daniele Lain\\
\affaddr{University of Padua}\\
\email{dlain@math.unipd.it}
\alignauthor Gene Tsudik\\
\affaddr{University of California, Irvine}\\
\email{gene.tsudik@uci.edu}
}

\maketitle

\begin{abstract}
Acoustic emanations of computer keyboards represent a serious privacy issue.
As demonstrated in prior work, physical
properties of keystroke sounds might reveal what a user is typing. However,
previous attacks assumed relatively strong adversary models that are not very
practical in many real-world settings. Such strong models assume: (i)
adversary's physical proximity to the victim, (ii) precise profiling of the
victim's typing style and keyboard, and/or (iii) significant amount of victim's
typed information (and its corresponding sounds) available to the adversary.

This paper presents and explores a new keyboard acoustic
eavesdropping attack that involves Voice-over-IP (VoIP), called \textit{Skype \& Type (\att)},
while avoiding prior strong adversary assumptions. This work
is motivated by the simple observation that people often engage in secondary
activities (including typing) while participating in VoIP calls. 
As expected, VoIP software acquires and faithfully
transmits all sounds, including emanations of pressed keystrokes, which 
can include passwords and other sensitive information. 
We show that one very popular VoIP software (Skype) conveys
enough audio information to reconstruct the victim's input -- keystrokes typed
on the remote keyboard. Our results demonstrate that, given some
knowledge on the victim's typing style and keyboard model, the attacker attains
top-5 accuracy of $91.7$\% in guessing a random key pressed by the victim.

Furthermore, we demonstrate that \att\ is
robust to various VoIP issues (e.g., Internet bandwidth fluctuations and 
presence of voice over keystrokes), thus confirming feasibility of this attack.
Finally, it applies to other popular VoIP software, such as Google Hangouts.
	
\end{abstract}

\section{Introduction}
Electronic devices are some of the most personal 
objects in many people's lives. We use them to store and manage private and sensitive information, such as 
photos, passwords, and messages. Protecting such sensitive data
by encryption is a common approach to prevent unauthorized access and
disclosure. However, there is no protection if data is leaked before 
encryption. In fact, eavesdropping on physical signals, such as acoustic or electromagnetic
emanations, is one way to recover either: (1) clear-text data before
encryption, e.g., during its input or visualization, or (2) encryption keys,
e.g., during data encryption and decryption. 
Indeed, the history of eavesdropping on physical signals dates back to 1943, 
when a Bell engineer discovered that an oscilloscope can retrieve the  
plain-text from electromagnetic emanations of a Bell Telephone model 131-B2 --
a mixing device used by the US Army to encrypt communications \cite{friedman1972}.

A common target for physical eavesdropping attacks are I/O
peripherals, such as keyboards, mice, touch-screens and printers. Examples of
prior physical eavesdropping attacks include: 
electromagnetic emanations of keyboards~\cite{Vuagnoux2009}, videos of users
typing on a keyboard~\cite{Balzarotti2008} or a touch-screen~\cite{shukla2014}, 
and keyboard acoustic emanations~\cite{Asonov2004}. The research community 
invested a lot of effort into studying keyboard acoustic emanations and demonstrated 
that it is a very serious privacy issue. A successful acoustic side-channel attack
allows an adversary to learn what a victim is typing, based on the sound produced by keystrokes.
Typically, sounds are recorded either directly, using microphones~\cite{Asonov2004,Halevi2012,Halevi2014,
Berger2006,Zhuang2009,Liu2015,Wang2014a,Zhu2014,martinasek2015},
or by exploiting various sensors (e.g., accelerometers~\cite{Marquardt2011,Wei2015}) to 
re-construct the same acoustic information. Once collected, the audio stream is typically analyzed using
techniques, such as supervised~\cite{Asonov2004,Halevi2012,Halevi2014,martinasek2015}
and unsupervised~\cite{Zhuang2009,Berger2006} machine learning, or
triangulation~\cite{Liu2015,Wang2014a,Zhu2014}. The final
result is a full or partial reconstruction of the victim's input.

It appears that all previous attacks require a 
compromised (i.e., controlled by the adversary) microphone near the victim's keyboard~\cite{Asonov2004,Halevi2012,Halevi2014,martinasek2015,Berger2006,Liu2015,Wang2014a,Zhu2014}. 
We believe that this requirement limits applicability of such attacks, thus reducing their real-world feasibility.
Although universal popularity of smartphones might ease 
placement of a compromised microphone (e.g., the one in the attacker's smartphone) 
close to the victim, the adversary still needs to either physically position and/or control it.
Moreover, some previous approaches are even more restrictive, requiring: (i) lots of training 
information to cluster~\cite{Berger2006}, thus necessitating long-term collection of keystroke 
sounds, or (ii) precise profiling of the victim's typing style and 
keyboard~\cite{Asonov2004,Halevi2012,Halevi2014,martinasek2015}.

\vfill\eject
In this paper, we present and explore a new keyboard acoustic
eavesdropping attack that: (1) does not require the adversary to control a microphone near
the victim, and (2) works with a limited amount of keystroke data. 
We call it \attackfull, or \attack for short\footnote{For more information and source code, please visit the project webpage: \url{http://spritz.math.unipd.it/projects/dst/}}. As a basis 
for this attack, we exploit Voice-over-IP (VoIP), one of the most popular and pervasive voice communication 
technologies used by great multitudes of people throughout the world. We premise
our work on a very simple observation and a hypothesis: 
\begin{quote} \small\sf
People involved in VoIP calls often engage in secondary activities, such as:
writing email, contributing their ``wisdom'' to social networks, reading news,
watching videos, and even writing research papers. Many of these activities
involve using the keyboard (e.g., entering a password). VoIP software automatically
acquires all acoustic emanations, including those of the keyboard, and transmits them 
to all other parties involved in the call. If one of these parties is malicious,
it can determine what the user typed based on keystroke sounds.
\end{quote}
We believe this work is both timely and important, especially, due to growing 
pervasiveness of VoIP software\footnote{In 2016, Skype reached $300$ million active
monthly users~\cite{skypeusers}.}.

Thus, remote keyboard acoustic eavesdropping attacks, if shown to be realistic, should concern every 
VoIP user. Prior studies \cite{Asonov2004,Halevi2012,Halevi2014,martinasek2015,Berger2006,Liu2015,Wang2014a,Zhu2014} 
have not considered either the setting of our attack, or the features of VoIP software. In particular, 
VoIP software performs a number of transformations on the sound before transmitting it over the
Internet, e.g., downsample, approximation, compression, and disruption of the stereo
information by mixing the sound into a single channel. Such transformations have not 
been considered in the past. In fact, for some prior results, these transformations conflict
with the assumptions, e.g., \cite{Liu2015,Wang2014a,Zhu2014} require 
stereo information for the recorded audio stream. 
Therefore, conclusions from these results are largely inapplicable to \attack.

\paragraph{Expected Contributions:}

\begin{compactitem}
\item We demonstrate \attack based on (remote) keyboard acoustic eavesdropping over VoIP software,
  with the goal of recovering text typed by the user during a VoIP call with the attacker. 
  \attack can also recover random text, such as randomly generated passwords or PINs. 
  We take advantage of spectral features of keystroke sounds and analyze them using
  supervised machine learning algorithms.
\item We evaluate \attack over a very popular VoIP software: {\bf Skype}. We designed a set of
  attack scenarios that we consider to be more realistic than those used in prior results on
  keyboard acoustic eavesdropping. 
  We show that \attack is highly accurate with minimal profiling of the victim's
  typing style and keyboard. It remains quite accurate even if neither profiling is
  available to the adversary. Our results show that \attack is very feasible,
  and applicable to real-world settings under realistic assumptions. It
  allows the adversary to recover, with high accuracy, typed (English) text, and to greatly 
  speed up brute-force cracking of random passwords. Moreover, preliminary experiments
  with Google Hangouts indicate that it is likely susceptible to \attack as well. 
\item We show, via extensive experiments, that \attack is robust to
  VoIP-related issues, such as limited available bandwidth that
  degrades call quality, as well as human speech over keystroke sounds.
\item Based on the insights from the design and evaluation phases of this work, 
we propose some tentative countermeasures to \att and similar
attacks that exploit spectral properties of keystroke sounds. 
\end{compactitem}

\paragraph{Organization.}
Section~\ref{sec:relwork} overviews related literature and state-of-the-art 
on keyboard eavesdropping. Next, Section~\ref{sec:sys} describes the system
model for our attack and various attack scenarios. Section~\ref{sec:attack}, presents 
\attack. Then, Section~\ref{sec:evaluation} evaluates \attack, 
discusses our results, the impact of VoIP-specific issues, and exhibits 
practical applications of \attack. Finally, Section~\ref{sec:countermeasures} proposes some
potential countermeasures, Section~\ref{sec:conclusions} summarizes
the paper and Section \ref{sec:future} overviews future work.

\section{Related Work}\label{sec:relwork}
Eavesdropping on keyboard input is an active and popular area of research. 
This section begins by overviewing attacks that rely strictly on acoustic emanations 
to recover the victim's typed text and then summarizes results that study
eavesdropping on other emanations, such as the WiFi signal, 
and surface vibrations.

However, there appears to be no prior research literature on taking advantage of 
acoustic emanations over the network, particularly over the Internet, to reconstruct
keyboard input --- which is instead the contribution of our work.

\paragraph{Attacks Using Sound Emanations.}\label{sec:relsound}
Research on keyboard acoustic eavesdropping started with the seminal paper of
Asonov and Agrawal~\cite{Asonov2004} who showed that, by training a neural network
on a specific keyboard, good performance can be achieved in eavesdropping on 
the input to the same keyboard, or keyboards of the same model. This work 
also investigated the reasons for this attack and discovered that the plate 
beneath the keyboard (where the keys hit the sensors) has a drum-like behavior. 
This causes the sound produced by different keys to be slightly distinct. 
Subsequent efforts can be divided based on whether
they use statistical properties of the sound spectrum or timing information. 

Approaches that use statistical properties of the spectrum typically 
apply machine learning, both supervised~\cite{Asonov2004,Halevi2012,Halevi2014,martinasek2015} 
and unsupervised~\cite{Berger2006,Zhuang2009} versions.  

Supervised learning techniques require many labeled samples and are highly
dependent on: (1) the specific keyboard used for training~\cite{Asonov2004}, and
(2) the typing style~\cite{Halevi2012,Halevi2014}. Such techniques use
Fast Fourier Transform (FFT) coefficients and neural networks to 
recover text that can also be random. Overall,
supervised learning approaches yield very high accuracy. However, this comes at 
the price of strong assumptions on how the data is collected: obtaining labeled 
samples of the acoustic emanations of the victim on his keyboard can be difficult
or unrealistic.

Unsupervised learning approaches can cluster together keys from sounds, or
generate sets of constraints between different key-presses. It is feasible 
to cluster key sounds and assign labels to the clusters by using relative letter  
frequency of the input language~\cite{Zhuang2009}. It is also possible to generate sets
of constraints from recorded sounds and select words from a dictionary
that match these constraints~\cite{Berger2006}. Unsupervised learning techniques 
have the advantage that they do not require ground truth. However, they make
strong assumptions on user input, such as obtaining many samples, i.e., 
emanations corresponding to a long text~\cite{Zhuang2009}, or requiring the
targets to be dictionary words~\cite{Berger2006}. They are less effective when 
keyboard input is random. 

An alternative approach involves analyzing timing information. One convenient way
to exploit timing information is using multiple microphones, such as the ones on mobile
phones~\cite{Liu2015,Wang2014a,Zhu2014}, and analyze the Time
Difference of Arrival (TDoA) information to triangulate the position of the
pressed key. Such techniques differ mostly in whether they require a training
phase~\cite{Wang2014a}, and rely on one~\cite{Liu2015} or more~\cite{Zhu2014} mobile
phones. 

\paragraph{Attacks Using Other Emanations.}\label{sec:relothers}
Another body of work focused on keyboard eavesdropping via non-acoustic
side-channels.

Typing on a keyboard causes its electrical components to emit electromagnetic waves, 
and it is possible to collect such waves, to recover the original keystrokes~\cite{Vuagnoux2009}. 
Furthermore, typing causes vibrations of the surface under the keyboard. These vibrations can 
be collected by an accelerometer (e.g., of a smartphone) and analyzed to determine 
pressed keys~\cite{Marquardt2011}.  

Analyzing movements of the user's hands and fingers on a
keyboard represents another way of recovering input. This is possible 
by video-recording a typing user~\cite{Balzarotti2008} or by 
using WiFi signal fluctuation on the user's laptop~\cite{Ali2015}.

\section{System and Threat models}\label{sec:sys}
To identify precise attack scenarios, we begin by defining the system
model that serves as the base for \att.
Section~\ref{sec:sysmodel} describes our assumptions 
about the victim and the attacker, and then carefully defines the problem of remote
keyboard acoustic eavesdropping. Section~\ref{sec:attacks} then presents some
realistic attack scenarios and discusses them in relation to the state-of-the-art.

\subsection{System Model}\label{sec:sysmodel}
\begin{figure}[htb]
	\centering
	\includegraphics[width=0.9\linewidth]{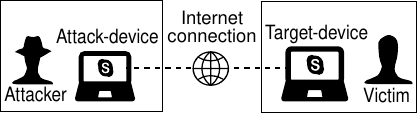}
	\caption[System model]{System model.}
	\label{fig:system}
\end{figure}

The system model is depicted in Figure~\ref{fig:system}.
We assume that the victim has a desktop or a laptop computer with a built-in
or attached keyboard, i.e., {\bf not} a smartphone or a tablet-like device.
Hereafter, it is referred to as \target. A genuine copy of
some VoIP software is assumed to be installed on \target; this software is not 
compromised in any way.  Also, \target\ is connected to the Internet and engaged 
in a VoIP call with at least one party who plays the role of the attacker.

\vspace{-0.5em}

The attacker is a malicious user who aims to learn some private information
about the victim. The attacker owns and fully controls a computer that we refer
to as \attdev, which has a genuine (unmodified) version of the same
VoIP software as \target. The attacker uses \attdev\ to receive and
record the victim's acoustic emanations using VoIP software. We assume 
that the attacker relies solely on information provided by VoIP software. In other words,
{\em during the attack}, the attacker receives no additional acoustic information from 
the victim, besides what VoIP software transmits to \attdev.

\subsection{Threat Model}\label{sec:attacks}
\attack\ transpires as follows: during a VoIP call between the victim and the attacker, the former 
types something on \target, e.g., a text of an email message or a password. We refer to this typed
information as  \tartxt. Typing \tartxt\ causes 
acoustic emanations from \target's keyboard, which are picked up by the \target's microphone 
and faithfully transmitted to \attdev\ by VoIP. The goal of the attacker is to learn \tartxt\
by taking advantage of these emanations.

We make the following assumptions:
\begin{compactitem}
\item As mentioned above, the attacker has no real-time audio-related information beyond 
  that provided by VoIP software. 
  Acoustic information can be degraded by VoIP software by downsampling and mixing. In
  particular, without loss of generality, we assume that audio is converted into a single (mono) signal, 
  as is actually the case with some VoIP software, such as Skype and Google Hangouts. 
\item If the victim discloses some keyboard acoustic emanations {\bf together}
  with the corresponding plaintext -- the actual pressed keys (called {\em ground truth}) --- 
  the volume of this information is small, on the order of a chat message or a short e-mail.
  We expect it to be no more than a few hundred characters.
\item \tartxt\ is very short (e.g., $\approx10$ characters) and random, corresponding to
 an ideal password. This keeps \attack as general as possible, since dictionary words are a ``special''
 case of random words, where optimization may be possible.
\end{compactitem}
We now consider some realistic \attack\ scenarios. We describe them starting with
the more generous setting where the attacker knows the victim's typing style and keyboard model, 
proceeding to the more challenging one where the attacker has neither type of information. 

\textsc{1) Complete Profiling:} 
In this scenario, the attacker knows some of the victim's keyboard acoustic 
emanations on \target, along with the ground truth for these emanations. This
might happen if the victim unwittingly provides some text samples to the attacker during the
VoIP call, e.g., sends chat messages, edits a shared document, or sends an
email message\footnote{Ground truth could also be collected offline, if the attacker happened
to be near the victim, at some point  before or after the actual attack. Note that this
still does not require physical proximity between the attacker and the victim in {\em real time}.}.
We refer to such disclosed emanations as \textit{``labeled data''}. To be realistic, 
the amount of labeled data should be limited to a few samples for each character. 

We refer to this as \textit{Complete Profiling} scenario, since the attacker
has maximum information about the victim. It
corresponds to attack scenarios used in prior supervised learning
approaches~\cite{Asonov2004,Halevi2012,Halevi2014,martinasek2015}, with the
difference that we collect acoustic emanations using VoIP software, while others 
collect emanations directly from microphones that are physically near \target.

\textsc{2) User Profiling:} In this scenario, we assume that the
attacker does not have any labeled data from the victim on \target.
However, the attacker can collect training data of the victim while
the victim is using the same type of device (including the keyboard) as
\target\footnote{In case the \targetfootnote is a desktop, knowing the model of
the desktop does not necessarily mean knowing the type of the
keyboard. However, in mixed video/audio call the keyboard model might be
visually determined, when the keyboard is placed in the visual range of the camera.}.
This can be achieved via social engineering techniques or with the help of an accomplice. 
We refer to this as \textit{User Profiling} scenario, since, unable to profile \target, 
the attacker profiles the victim's typing style on the same device type.

\textsc{3) Model Profiling:} This is the most challenging, though
the most realistic, scenario. The attacker has absolutely no 
training data for the victim.The attacker and the victim
are engaged in a VoIP call and information that the attacker 
obtains is limited to victim keyboard's acoustic emanations.

The attacker's initial goal is to determine what laptop the victim is
using. To do so, we assume that the attacker maintains a database of
sounds from previous attacks. If the attacker already
profiled the model of the current victim's \target, it can use this information  
to mount the attack. We refer to this as \textit{Model Profiling} scenario, since 
although the attacker can not profile the current victim, it can still profile a device
of the same model as \target.

\section{Skype \& Type attack}\label{sec:attack}
\begin{figure}[htb]
	\centering
	\includegraphics[width=0.95\linewidth]{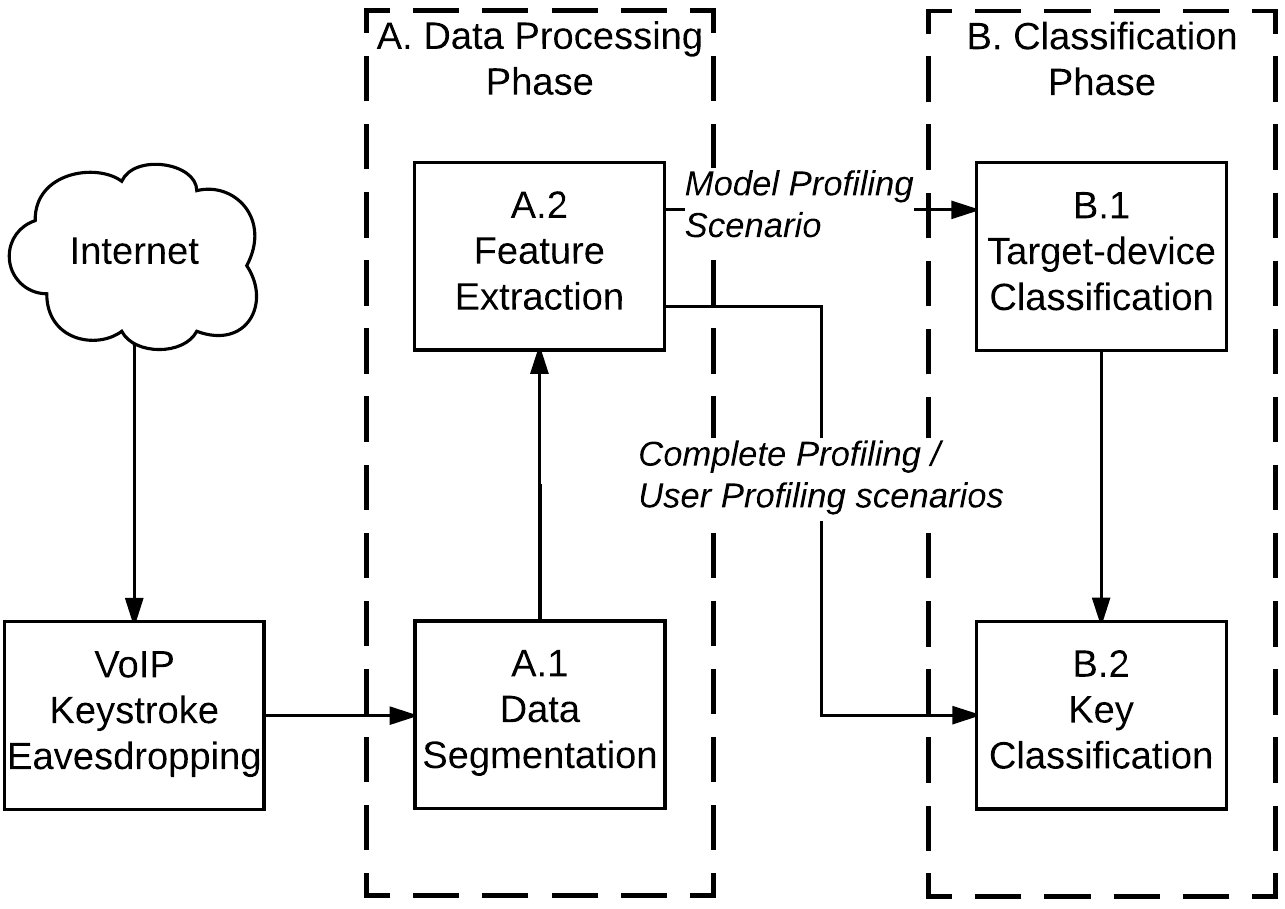}
	\caption[Overview of \attack]{\attack\ steps.} 
	\label{fig:attack}
\end{figure}

This section provides a detailed description of \attack. Recall that all envisaged
scenarios involve the attacker engaged in a VoIP call with the victim.
During the call, the victim types something on \target's keyboard. 
\attack\ proceeds as described below and illustrated in Figure~\ref{fig:attack}.

\vspace{-0.5em}

First, the attacker receives and records acoustic emanations of \target's
keyboard over VoIP. One way to do so is by channeling VoIP output to 
some local recording software. Then, the actual attack involves two 
phases: (i) data processing, and (ii) data classification. Each phase involves two steps:
\begin{compactenum}
\item Data processing includes data segmentation and feature extraction steps.
They are performed in each of the three attack scenarios defined in Section~\ref{sec:sys}. 
\item Data classification phase includes \target\ classification and key classification steps.
Their execution depends on the specific attack scenario:

-- In \textit{Complete Profiling} and \textit{User Profiling} scenarios, the
  attacker already profiled the victim, either on \target\ 
  (\textit{Complete Profiling}) or on a device of the same model
  (\textit{User Profiling}). The attacker uses this data as a training set,
  and proceeds to classify \tartxt. This case is indicated in Figure~\ref{fig:attack} by the path 
  where key classification follows feature extraction.
  
-- In \textit{Model Profiling} scenario, since the attacker has no knowledge of the
  victim's typing style or \target, it begins by trying to identify 
  \target\ by classifying its keyboard sounds. The attacker then proceeds
  to classify \tartxt\ by using correct training data. This case is
  indicated in Figure~\ref{fig:attack} by the path where \target\
  classification is the next step after feature extraction.
\end{compactenum}
Next, we describe these two phases in more detail.

\subsection{Data Processing Phase}\label{subsec:features}
The main goal in this phase is to extract meaningful features
from acoustic information. The first step is 
{\em data segmentation} needed to isolate distinct keystroke sounds within the
recording. Subsequently, using these sound samples, we build derived values
(called features) that represent properties of acoustic information. This
step is commonly referred to as {\em feature extraction}.

\subsubsection{Data Segmentation}
\begin{figure}[htb]
	\centering
	\includegraphics[width=0.95\linewidth]{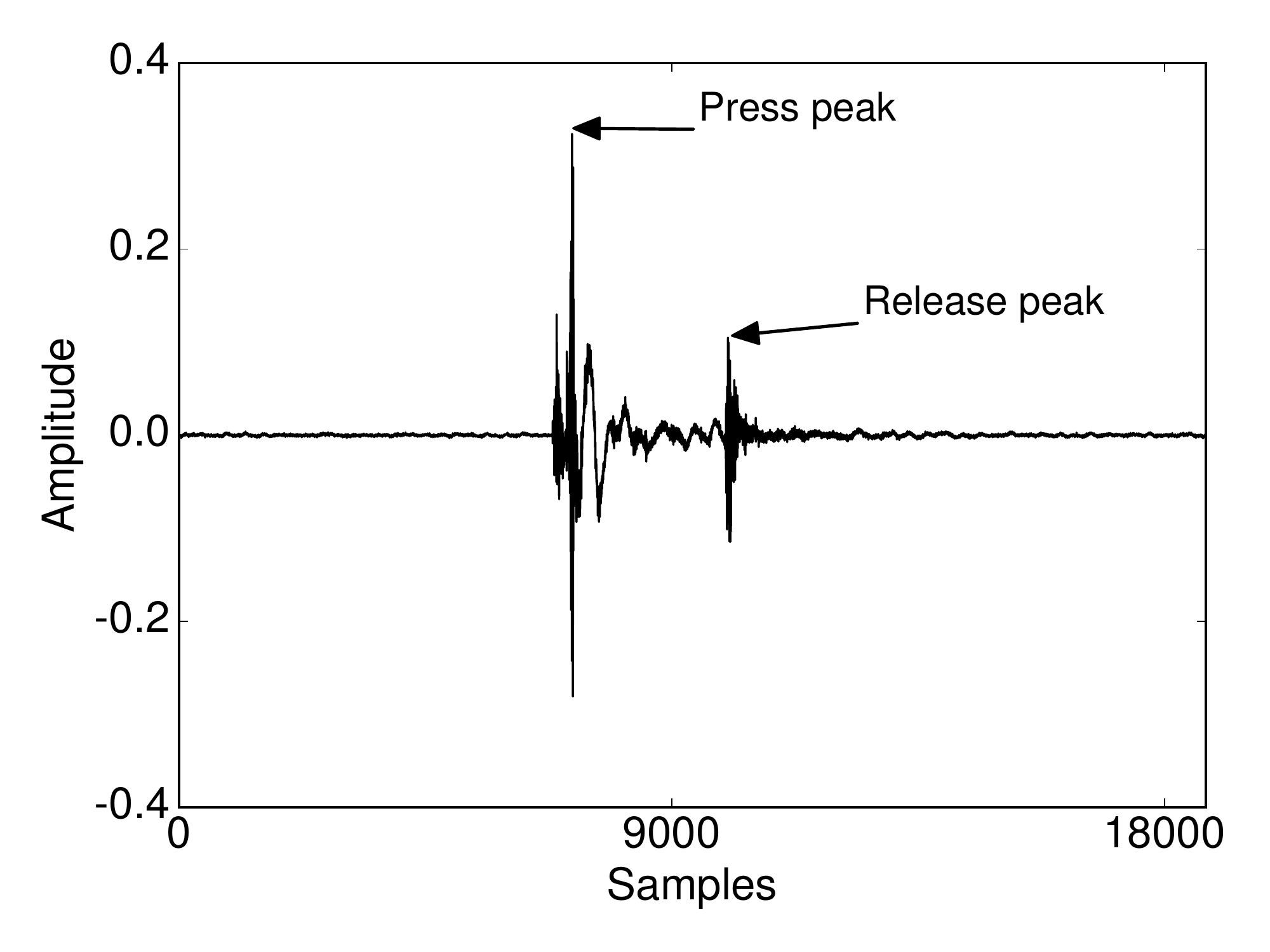}
	\caption[Waveform of a keystroke]{Waveform of the ``\textit{A}'' key, 
		recorded on an Apple Macbook Pro 13" laptop.}
	\label{fig:waveform}
\end{figure}

We perform data segmentation according to the following
observation: the waveform of a keystroke sound presents two distinct
peaks, shown in Figure~\ref{fig:waveform}.
These two peaks correspond to the events of: (1) the finger pressing the key --
\textit{press} peak, and (2) the finger releasing the key -- \textit{release}
peak. Similar to~\cite{Asonov2004}, we only use the press
peak to segment the data and ignore the release peak. This is because the former 
is generally louder than the latter and is thus easier to isolate,
even in very noisy scenarios.

\vspace{-0.5em}

To perform automatic isolation of keystrokes, we set up a detection mechanism as
follows: we first normalize the amplitude of the signal to have root mean square
of $1$. We then sum up the FFT coefficients over small windows of $10$ms, to
obtain the energy of each window. We detect a press event when the energy of a
window is above a certain threshold, which is a tunable parameter. We then extract the subsequent $100$ms~\cite{Berger2006,Zhuang2009} as the
waveform of a given keystroke event. If sounds of pressed keys are very closely spaced,
it is possible to extract a shorter waveform.

\subsubsection{Feature Extraction} 
As features, we extract the mel-frequency cepstral coefficients (MFCC)~\cite{logan2000}. 
These features capture statistical properties of the sound spectrum, which is the only
information that we can use. Indeed, due to the mono acoustic information,
it is impossible to set up an attack that requires stereo audio and uses
TDoA, such as~\cite{Liu2015,Wang2014a,Zhu2014}. Among possible
statistical properties of the sound spectrum -- including: MFCC, FFT
coefficients, and cepstral coefficients -- we chose MFCC which yielded the best
results. To select the most suitable property, we ran the following experiment: 
\begin{quote}
Using a Logistic Regression classifier we classified a dataset with $10$ samples 
for each of the $26$ keys corresponding to the letters of the English alphabet, 
in a $10$-fold cross-validation scheme. We then evaluated the accuracy of the classifier with 
various spectral features: FFT coefficients, cepstral coefficients, and MFCC.  
\end{quote}
We repeated this experiment with data from five users on a Macbook Pro laptop.
Accuracy results were: 90.61\% ($\pm$ 3.55\%) for MFCC, 86.30\% ($\pm$ 6.34\%) for FFT
coefficients, and 51\% ($\pm$ 18.15\%) for cepstral coefficients. This shows that 
MFCC offers the best features.
For MFCC experiments we used parameters similar to those in~\cite{Zhuang2009}:
a sliding window of $10$ms with a step size of $2.5$ms, $32$ filters in the mel scale 
filterbank, and used the first $32$ MFCC.

\subsection{Classification Phase}\label{subsec:classifiers}
In this phase, we apply a machine learning algorithm to features extracted in
the Data Processing phase, in order to perform:
\begin{compactitem}
\item {\Target classification} using all keystroke sound emanations 
  that the attacker received.
\item {Key classification} of each single keyboard key of \target, by using sound 
 emanations of the keystrokes.
\end{compactitem}
Each classification task is performed depending on the scenario. In
\textit{Complete Profiling} and \textit{User Profiling} scenarios,
the attacker already profiled the victim on \target, or on a device of the same model, 
respectively. Then, the attacker loads correct training data and performs
the key classification task, to understand \tartxt. 

In contrast, in {\em Model Profiling} scenario, the attacker first performs \target\ classification
task, in order to identify the model. Next, the attacker loads correct training data,
and proceeds to the key classification task.

The only viable machine learning approach for both the key and \target\
classification tasks is a supervised learning technique. As
discussed in Section~\ref{sec:attacks}, approaches that require lots of data to
cluster, such as \cite{Berger2006}, are incompatible with our
assumptions, because we might have only a small amount of both training and testing
data. Moreover, potential randomness of \tartxt\ makes it impossible to realize
constraint-based approaches, which would require \tartxt\ to be a
meaningful word, as in \cite{Zhuang2009}.

\subsubsection{Target-device Classification}\label{sec:lapclass}
We consider the task of \target\ classification as a multiclass classification
problem, where different classes correspond to different \target\
models known to the attacker. More formally, we define the problem as follows: 
\begin{quote}
We start with a number of samples $s\in~S$, each 
represented by its feature vector $\vec{s}$, and generated by the same \target\
$l$ of model $\tilde{l}$, among a set $\mathcal{L}$ of known \target\ models. We
want to know which \target\ model generated the samples in $S$, by classifying
every sample $s$, and then taking the mode of these predictions.
\end{quote}
To perform this classification task, we use a $k$-nearest neighbors ($k$-NN) 
classifier with $k=10$ neighbors, that outperformed other classifiers such as 
Random Forest and Logistic Regression in our preliminary experiments.

\subsubsection{Key Classification} \label{sec:keyclass}
We consider key classification to be a multiclass classification
problem, where different classes correspond to different keyboard keys. 
To evaluate the classifier's quality we use
\textit{accuracy} and \textit{top-n accuracy} measures. Given true
values of $k$, accuracy is defined in the multiclass classification case
as the fraction of correctly classified samples over all samples. 
Top-n accuracy is defined similarly. The sample is correctly classified if it is
present among the top $n$ guesses of the classifier.

To perform key classification, we use a Logistic Regression (LR)
classifier, since it outperformed all others, including: Linear
Discriminant Analysis (LDA), Support Vector Machines (SVM), 
Random Forest (RF), and $k$-nearest neighbors. We show this in an
experiment which uses each candidate to classify a
dataset of $10$ samples, for each of the $26$ keys corresponding to the
letters of the English alphabet, in a $10$-fold cross-validation scenario. We use
MFCC as features, and, for each classifier, we optimize the hyper-parameters
with an extensive grid search.

\begin{figure}[htb]
	\includegraphics[width=0.95\linewidth]{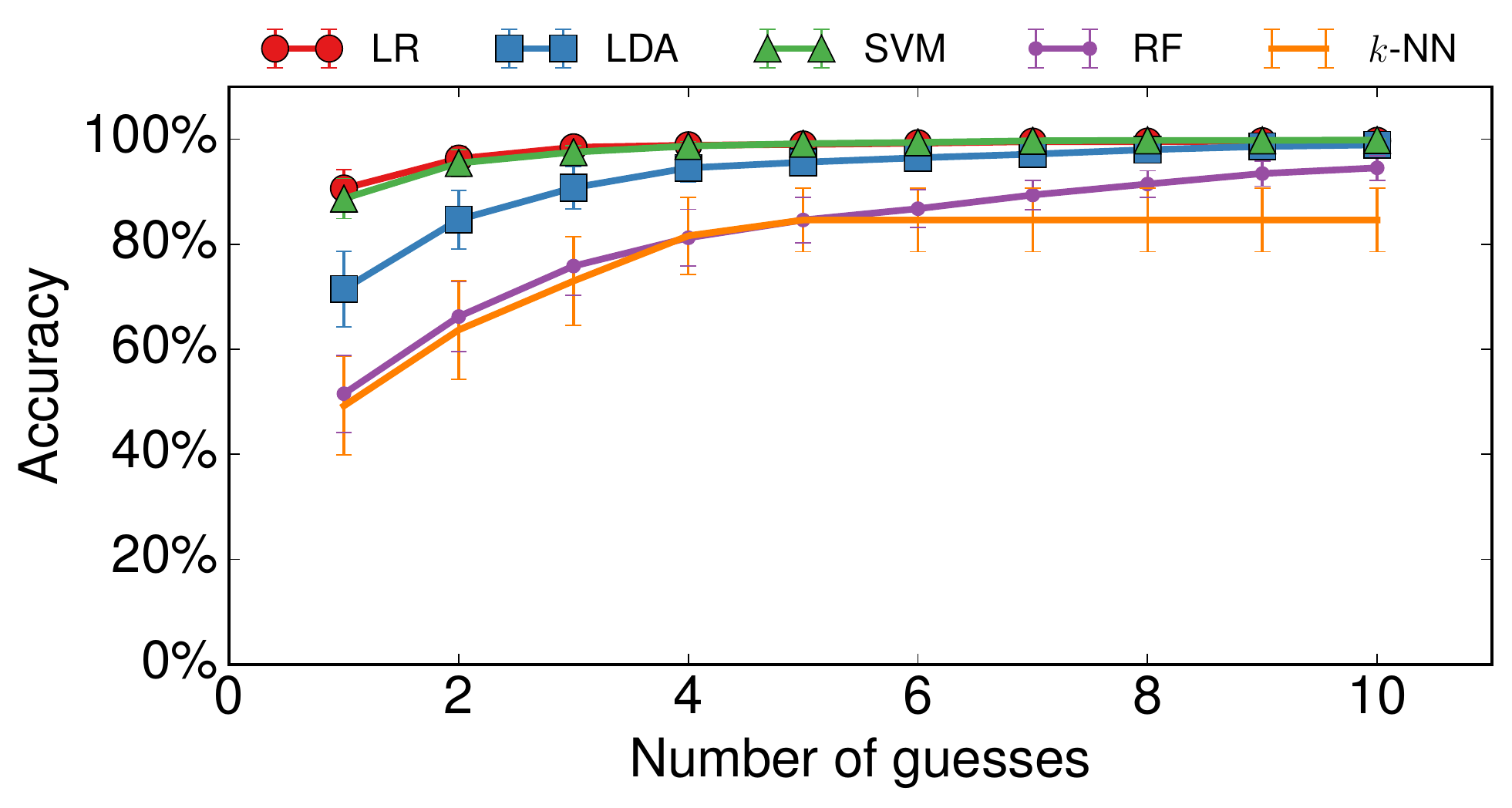}
	\caption[Performance of various classifiers]{Average top-n accuracy of single key
		classification, as a function of the number of guesses, for each classifier.}
	\label{fig:classifiers}
\end{figure}

Results are shown in Figure~\ref{fig:classifiers} which demonstrates that 
the best performing classifiers are LR and SVM. This is especially the case if the classifier is allowed to 
make a small number of predictions (between $1$ and $5$), which is more realistic 
in an eavesdropping setting. In particular, both LR and SVM exhibit around $90$\% top-1
accuracy, and over $98.9$\% top-5 accuracy. However, LR slightly outperforms SVM until top-4.

\section{Evaluation}\label{sec:evaluation}
To assess feasibility of \attack\ we conducted a set of experiments
that cover all previously described scenarios. We chose {\bf Skype} as the 
underlying VoIP software. There are three reasons for this choice: (i) Skype is
one of the most popular VoIP tools~\cite{skypeusers,skypeminutes,skypemobile};
(ii) its codecs are used in Opus, an IETF standard~\cite{valin2012},
employed in many other VoIP applications, such as Google Hangouts and
Teamspeak~\cite{opussupport}; (iii) it reflects our general assumption about 
mono audio. Therefore, we believe Skype is representative of a wide range of
VoIP software packages and its world-wide popularity makes it appealing for
attackers. Even though our \att\ evaluation is focused on Skype,
preliminary results show that we could obtain similar results with other VoIP software,
such as Google Hangouts.

We first describe experimental data collection in Section~\ref{subsec:datasets}. 
Then, we discuss experimental results  in Section~\ref{subsec:attacks_eval}. 
Next, Section~\ref{sec:noise}) considers several issues in using VoIP and 
Skype to perform \attack, e.g., impact of bandwidth reduction on audio quality, and 
the likelihood of keystroke sounds overlapping with the victim's voice. 
Finally, in Section~\ref{subsec:applications}, we report on \attack\ results in the context 
of two practical scenarios: understanding English words, and improving brute-force cracking of random 
passwords.

\subsection{Data Collection}\label{subsec:datasets}
\begin{figure}[htb]
	\centering
	\includegraphics[width=0.95\linewidth]{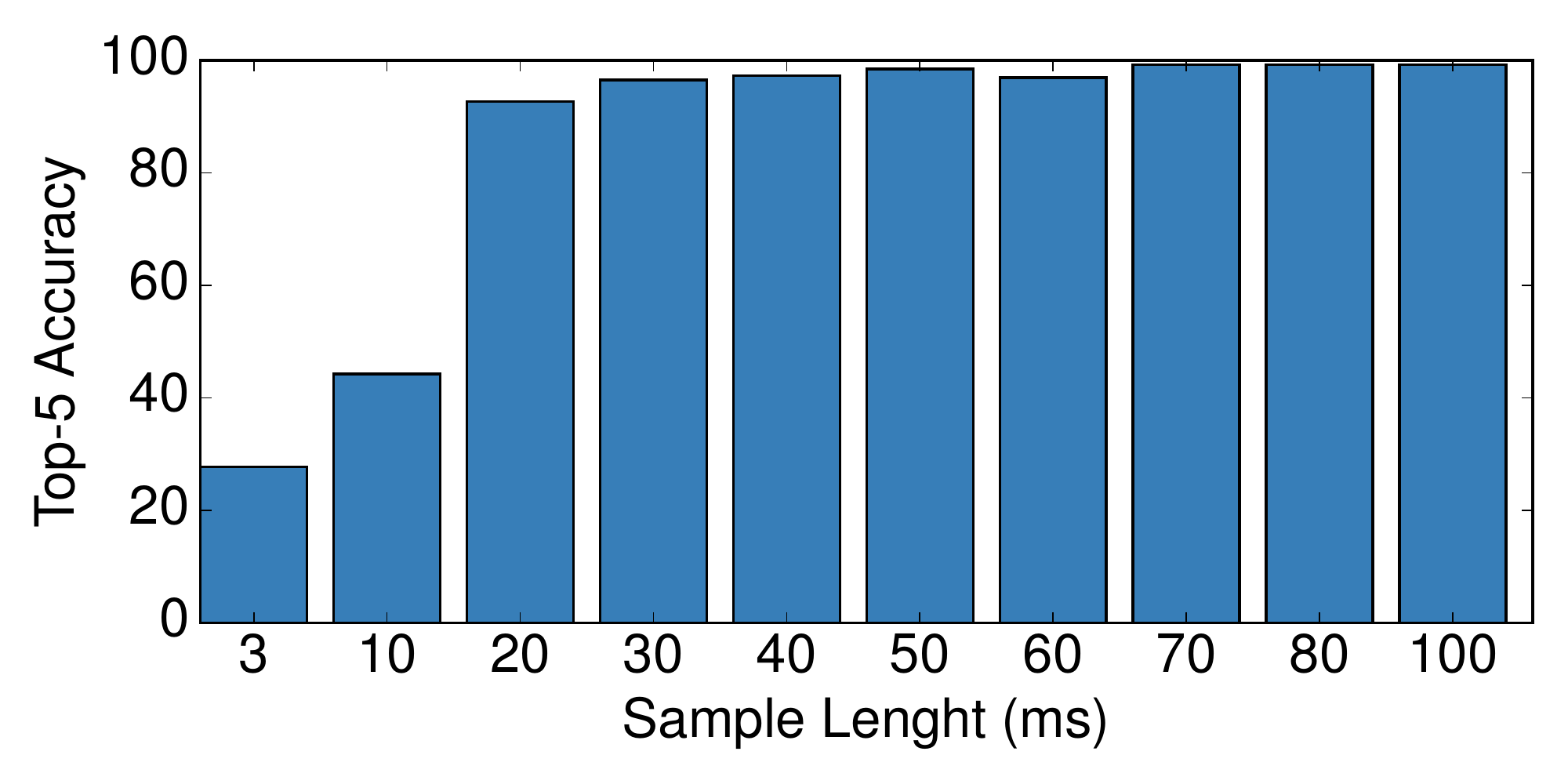}
	\caption{Top-5 accuracy of single key classification for different sample lengths.}
	\label{fig:window_size}
\end{figure}

We collected data from five distinct users. For each  user, the task was to press
the keys corresponding to the English alphabet, sequentially from ``A'' to
``Z'', and to repeat the sequence ten times, first by only using the right index
finger (this is known as \textit{Hunt and Peck} typing, referred to as \textit{HP}
from here on), and then by using all fingers of both hands (\textit{Touch}
typing)~\cite{Halevi2014}. We believe that typing letters in the order
of the English alphabet rather than, for example, typing English words, did not
introduce bias. Typing the English alphabet in order is similar to typing random 
text, that \attack targets. Moreover, a fast touch typist usually takes around $80$ms
to type consecutive letters \cite{card1980keystroke}, and \attack works without any accuracy 
loss with samples shorter than this interval. In order to test correctness of this assumption,
we ran a preliminary experiment as follows:
\begin{quote}
We recorded keystroke audio of a single user on a Macbook Pro laptop 
typing the English alphabet sequentially from ``A'' to ``Z'' via Touch typing. 
We then extracted the waveforms of the letters, as described 
in Section~\ref{subsec:features}. However, instead of extracting 100ms of the waveform, 
we extracted 3ms~\cite{Asonov2004}, and from 10ms to 100ms at intervals of 10ms for each step. 
We then extracted MFCC and tested \attack in a $10$-fold cross-validation scheme. Figure \ref{fig:window_size} shows top-5 accuracy of this preliminary experiment, for different
lengths of the sound sample that we extracted. 
\end{quote}
We observe that, even with very short 20ms samples, \attack suffers minimal accuracy loss. 
Therefore, we believe that adjacent letters do not influence each other, since sound 
overlapping is very unlikely to occur.

Note that collecting only the sounds corresponding to letter keys, instead of those
for the entire keyboard, does not affect our
experiment. The ``acoustic fingerprint'' of every key is related to its
position on the keyboard plate~\cite{Asonov2004}.  Therefore, all keys
behave, and are detectable, in the same way~\cite{Asonov2004}.  Due to this
property, we believe that considering only letters is sufficient to prove our
point. Moreover, because of this property, it would be trivial to extend our approach 
to various keyboard layouts, by associating the keystroke sound with
the position of the key, rather than the symbol of the key, and then mapping the
positions to different keyboard layouts.

Every user ran the experiment on six laptops: (1) two Apple Macbooks Pro 13'' mid
2014, (2) two Lenovo Thinkpads E540, and (3) two Toshiba Tecras M2. We selected these
as being representative of many common modern laptop models: 
Macbook Pro is a very popular aluminium-case high-end laptop, Lenovo
Thinkpad E540 is a 15'' mid-priced laptop, and Toshiba Tecra M2 is an older
laptop model, manufactured in 2004. All acoustic emanations of the laptop
keyboards were recorded by the microphone of the laptop in use, with
Audacity software v2.0.0. We recorded all data with a sampling frequency of $44.1$kHz, 
and then saved it in WAV format, $32$-bit PCM signed.

We then filtered the results by routing the recorded emanations through the Skype software,
and recording the received emanations on a different computer (i.e., on the attacker's side).
To do so, we used
two machines running Linux, with Skype v4.3.0.37, connected to a
high-speed network. 
During the calls, there was no sensible data loss. We analyzed 
bandwidth requirements needed for data loss to occur, and the impact of
bandwidth reduction, in Section~\ref{subsec:skype}. 

At the end of data collection and processing phases, we obtained
datasets for all the five users on all six laptops, with both the HP and
Touch-typing styles. All datasets are both unfiltered, i.e., raw
recordings from the laptop's microphone, and filtered through Skype and recorded on the attacker's machine. Each
dataset consists of 260 samples, 10 for each of the 26 letters of the English
alphabet. The number of users and of laptops we considered often exceeds
related work on the topic~\cite{Asonov2004,Halevi2012,Halevi2014,martinasek2015}, 
where only a maximum of 3 keyboards were tested, and a single test user.

\subsection{S\&T Attack Evaluation}\label{subsec:attacks_eval}
We evaluated \attack with all scenarios described in Section~\ref{sec:attacks}. We evaluated
\textit{Complete Profiling} scenario in detail, by analyzing 
performance of \attack separately for all three laptop models, two
different typing styles, and VoIP filtered and unfiltered data. We consider
this to be a favorable scenario for showing the accuracy of \attack. In
particular, we evaluated performance by considering VoIP
transformation, and various combinations of laptops and typing styles. We then
analyzed only the realistic combination of Touch typing data,
filtered with Skype.

We evaluated \attack accuracy in recognizing single characters, according
to the top-n accuracy, defined in~\cite{boyd2012accuracy}, as mentioned in 
Section \ref{sec:keyclass}. As a baseline, we considered a random guess with
accuracy $\frac{x}{l}$, where $x$ is the number of guesses, and $l$ is the size
of the alphabet. Therefore, in our experimental setup, accuracy of the
random guess is $\frac{x}{26}$, since we considered 26 letters of the English
alphabet. Because of the need to eavesdrop on random text, we can not use
``smarter'' random guesses that, for example, take into account 
letter frequencies in a given language.

\subsubsection{Complete Profiling Scenario}\label{sec:atk1}
To evaluate the scenario where the victim disclosed some labeled data to the
attacker, we proceeded as follows. We considered all datasets, one at a time,
each consisting of 260 samples (10 for every letter), in a stratified 10-fold 
cross-validation scheme\footnote{In a
stratified $k$-fold cross-validation scheme, the dataset is split in $k$ sub-samples 
of equal size, each having the same percentage of samples for every class
as the complete dataset.
One sub-sample is used as testing data, and the other $(k-1)$ -- as
training data. The process is repeated $k$ times, using each of the sub-samples as 
testing data.}. For every fold, we performed feature selection on training data 
using a Recursive Feature Elimination algorithm~\cite{guyon2002}. We calculated the 
classifier's accuracy over each fold, and then computed
the mean and standard deviation of accuracy values.

\begin{figure}[htb]
	\centering
	\includegraphics[width=0.95\linewidth]{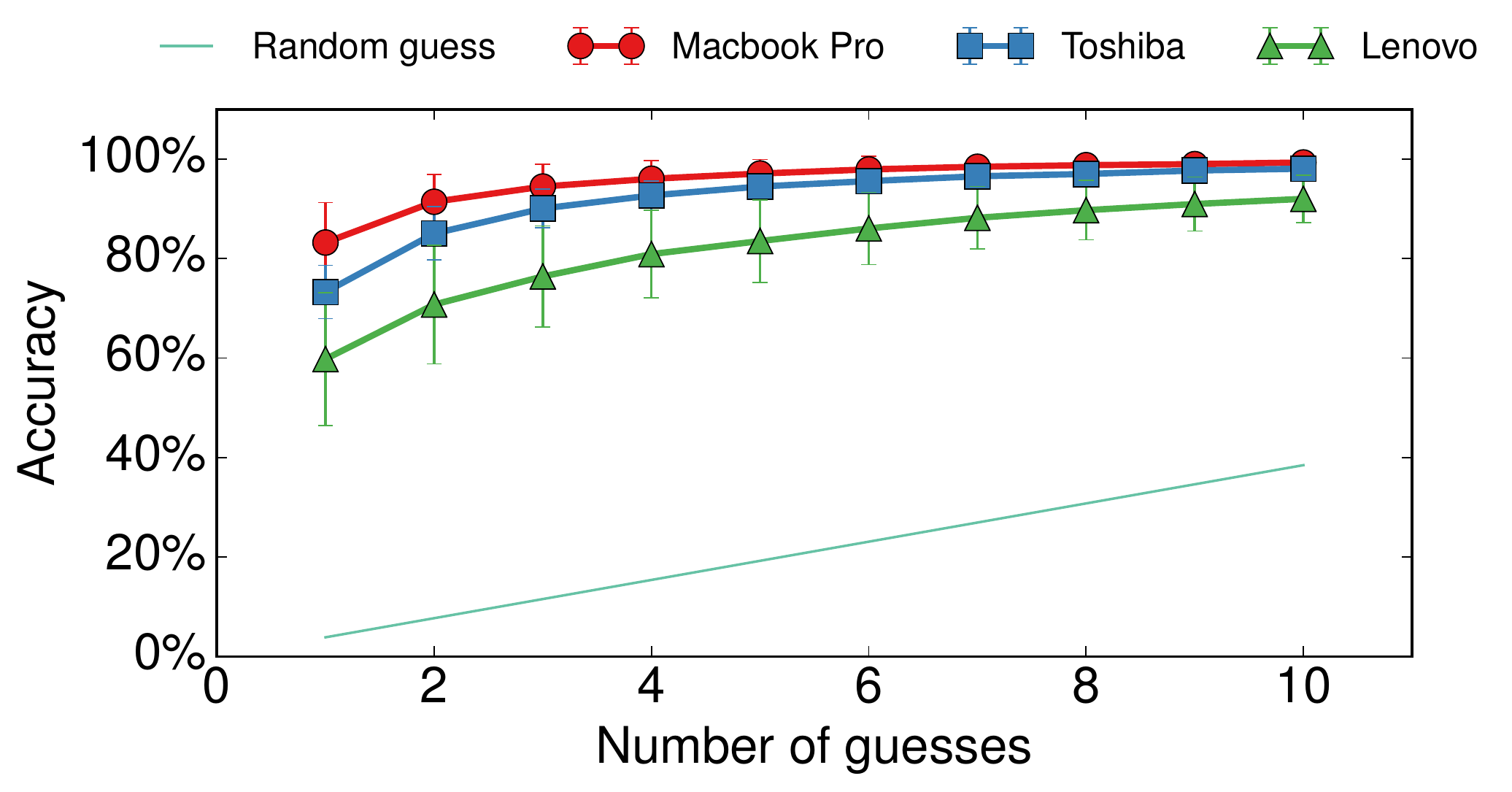}
	\caption{\attack performance -- \textit{Complete Profiling} scenario, Touch typing, 
	Skype-filtered data, average accuracy.}
	\label{fig:zero_touch_full}
\end{figure}

Figure~\ref{fig:zero_touch_full} depicts results of the experiment on the realistic
Touch typing, Skype-filtered data combination. We observe that \attack achieves its
lowest performance on Lenovo
laptops with top-1 accuracy of 59.8\%, and a top-5 accuracy 83.5\%. On
Macbook Pro and Toshiba, we obtained a very high top-1 accuracy,
83.23\% and 73.3\% respectively, and a top-5 accuracy of 97.1\% and 94.5\%,
respectively. We believe that these differences
are due to variable quality of manufacturing, e.g., the keyboard of
our particular Lenovo laptop model is made of cheap plastic materials.

Interestingly, we found that there is little difference between this data combination
(that we consider the most unfavorable) and the others. In particular,
we compared average accuracy of \attack on HP and Touch typing data, and found that
the average difference in accuracy is 0.80\%. Moreover, we compared the results
of unfiltered data with Skype filtered data, and found that the average difference in accuracy
is a surprising 0.33\%. This clearly shows that Skype does not reduce accuracy
of \attack. 

We also ran a smaller set of these experiments over Google Hangouts 
and observed the same tendency. This means that the keyboard acoustic 
eavesdropping attack is applicable to other VoIP software, not only Skype.
It also makes this attack more credible as a real threat. We report these results in more detail in Appendix \ref{sec:appendix1}.

From now on, we only focus on the most realistic combination -- Touch typing and Skype filtered 
data. We consider this combination to
be the most realistic, because \attack is conducted over Skype, and it is
more common for users to type with the Touch typing style, rather than the HP
typing style. We limit ourself to this combination to further understand
real-world performance of \attack.

\subsubsection{A More Realistic Small Training Set}\label{sec:atk1bis}
\begin{figure}[h!]
	\centering
	\includegraphics[width=0.95\linewidth]{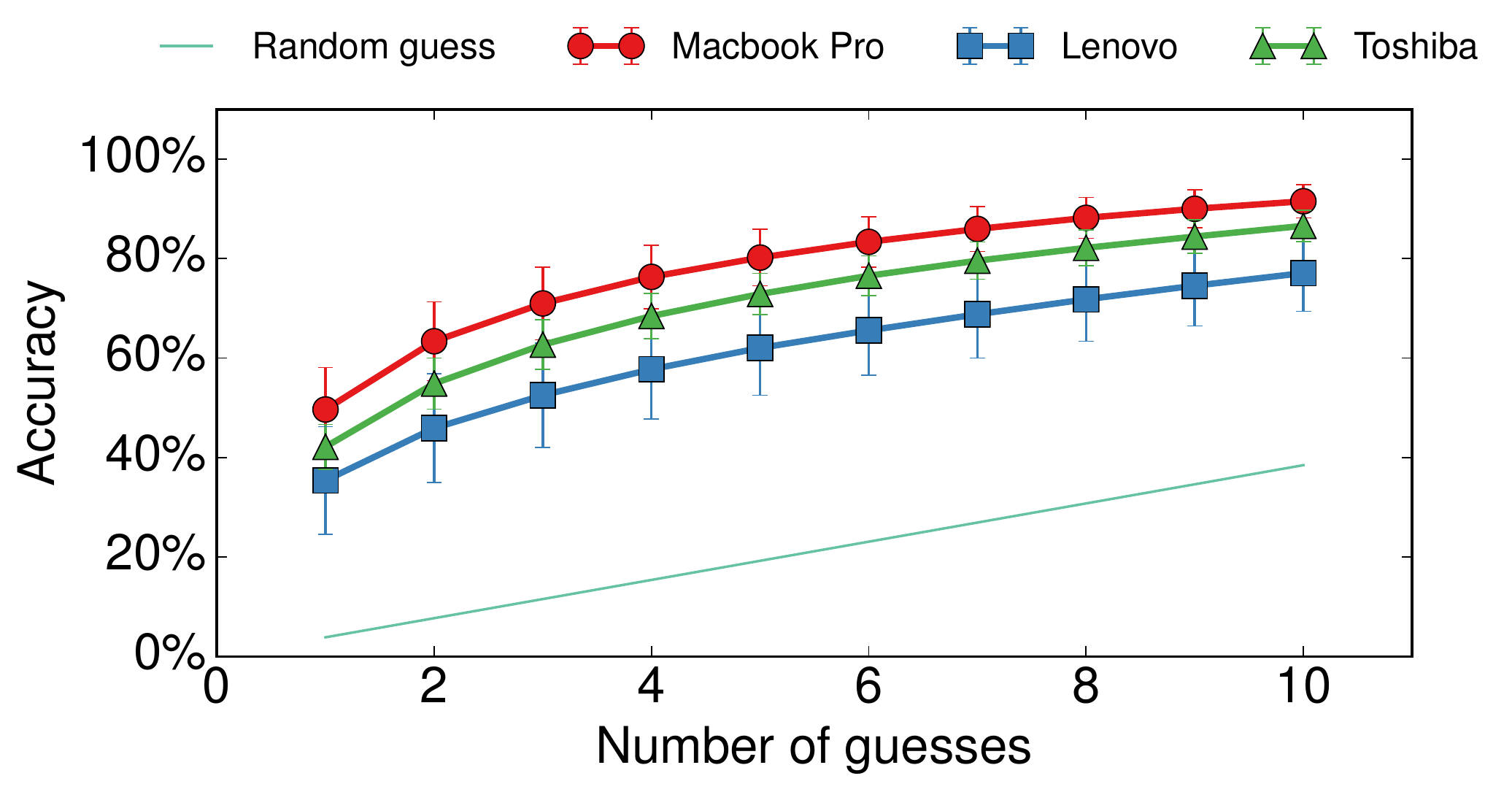}
	\caption[Attack accuracy with a small training set]{\attack performance -- \textit{Complete Profiling}
		scenario, average accuracy, on a small subset of 105 samples that respects the
		letter frequency of the English language.}
	\label{fig:small_train}
\end{figure}

As discussed in Section~\ref{sec:attacks}, one way to mount \attack\ in 
the \textit{Complete Profiling} scenario is by exploiting data accidentally
disclosed by the victim, e.g., via Skype instant-messaging with the attacker 
during the call. However, each  dataset we collected includes 10 repetitions
of every letter,  from ``A'' to ``Z'', 260 total. Though this is a reasonably
low amount, it has unrealistic letter frequencies. We therefore trained the
classifier with a small subset of  training data that conforms to the letter
frequency of the English language. To do this, we retained 10 samples of the
most frequent letters according to the Oxford
Dictionary~\cite{oxfordletters}. Then, we randomly excluded samples of less
frequent letters until only one sample for the least frequent letters was
available. Ultimately, the subset contained 105 samples, that might correspond to
a typical short chat message or a brief email. We then evaluated 
performance of the classifier trained with this subset, on a 10-fold
cross-validation scheme. This random exclusion scheme was repeated 20 times for
every fold. Results on Touch typing Skype filtered data are shown in
Figure~\ref{fig:small_train}.

We incurred an accuracy loss of around 30\% on every laptop, mainly
because the (less frequent) letters for which we have only a few examples in the
training set are harder to classify. However, performance of the
classifier is still good enough, even with such a very small training set, composed
of 105 samples with realistic letter frequency. This further motivates the
\textit{Complete Profiling} scenario: the attacker can exploit even a few
acoustic emanations that the victim discloses via a short message 
during a Skype call.

\subsubsection{User Profiling Scenario}\label{sec:atk2}
In this case, the attacker profiles the victim on a laptop of the
same model of \target.  We selected the dataset of a particular user on one of the six
laptops, and used it as our training set. Recall that it
includes 260 samples, 10 for every letter. This training set 
modeled data that the attacker acquired, e.g., via social engineering
techniques. We used the dataset of the same user on the other laptop of the same
type, to model \target. We conducted this experiment for
all six laptops.

\begin{figure}[h!]
	\centering
	\includegraphics[width=0.95\linewidth]{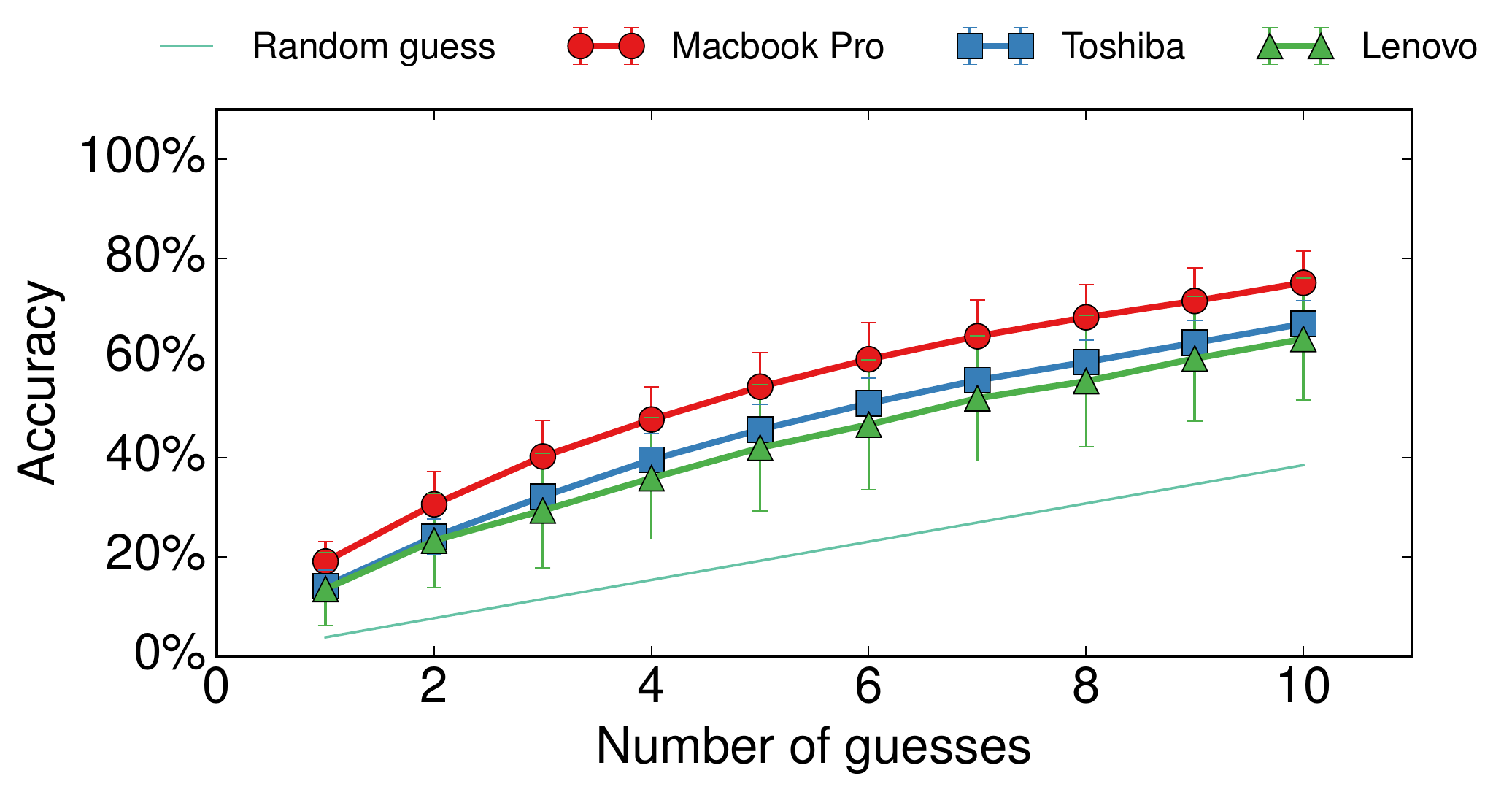}
	\caption[Attack accuracy, \textit{User Profiling} scenario]{\attack performance -- \textit{User
			Profiling} scenario, average accuracy.}\label{fig:one_all}
\end{figure}

Results reflected in Figure~\ref{fig:one_all} show that top-1 accuracy decreases to as low as 
14\% on Toshiba and Lenovo laptops, and to 19\% on Macbook Pro. However, 
top-5 accuracy grows to 41.9\%, 54\%, and 45.6\% on Lenovo, Macbook Pro, and Toshiba,
respectively. This shows the utility of social engineering techniques used to obtain labeled 
data of the victim, even on a different laptop. 

\subsubsection{Model Profiling Scenario}\label{sec:atk3}
We now evaluate the most unfavorable and the most realistic scenario where the
attacker does not know anything about the victim. Conducting
\attack in this scenario requires: (i) \target\ classification, followed by (ii) key classification. 

\paragraph{\Targetbf classification.}\label{sec:atk31}
The first step for the attacker is to determine whether \target is a
known model. We assume that the attacker collected a database of acoustic
emanations from many keyboards. 

When acoustic emanations from \target\ are
received, if the model of \target is present in the database, the attacker can 
use this data to train the classifier. To evaluate this scenario, we completely 
excluded all records of one user and of one
specific laptop of the original dataset. We did this to create a training set
where both the victim's typing style and the victim's \target are unknown to the
attacker. We also added, to the training set, several devices,
including 3 keyboards: Apple Pro, Logitech Internet, Logitech Y,
as well as 2 laptops: Acer E15 and Sony Vaio Pro 2013. 

We did this to show that a laptop is recognizable from its keyboard acoustic
emanations among many different models. We evaluated the accuracy of $k$-NN
classifier in idenitifying the correct laptop model, on the Touch typing and
Skype filtered data combination. Results show quite high accuracy of 93\%. 
This experiment confirms that an attacker can determine the victim's device,
by using acoustic emanations.

We now consider the case when the model of \target is not in the database.
The attacker must first determine that this is indeed so. This can be done
using the confidence of the classifier. If \target is in the database, most 
samples are classified correctly, i.e., they ``vote''correctly. However, when \target\
is not in the database, predicted labels for the samples are spread among known 
models. One way to assess whether this is the case is to calculate the
difference between the mean and the most-voted labels. We observed that trying to
classify an unknown laptop consistently leads to a lower value of this metric: 0.21 {\em vs}
0.45. The attacker can use such observations, and then attempt to obtain further
information via social engineering techniques, e.g.,  
laptop~\cite{kohno2005}, microphone~\cite{das2014} or webcam~\cite{lukas2006}
fingerprinting. 

\paragraph{Key classification.}\label{sec:atk32}
Once the attacker learns \target, it proceeds to determine keyboard input. 
However, it does not have any extra information about the 
victim that can be used to train the classifier. Nonetheless, the attacker 
can use, as a training set, data obtained from another user on a laptop of the same model 
as \target.

\begin{figure}[htb]
	\centering
	\includegraphics[width=0.95\linewidth]{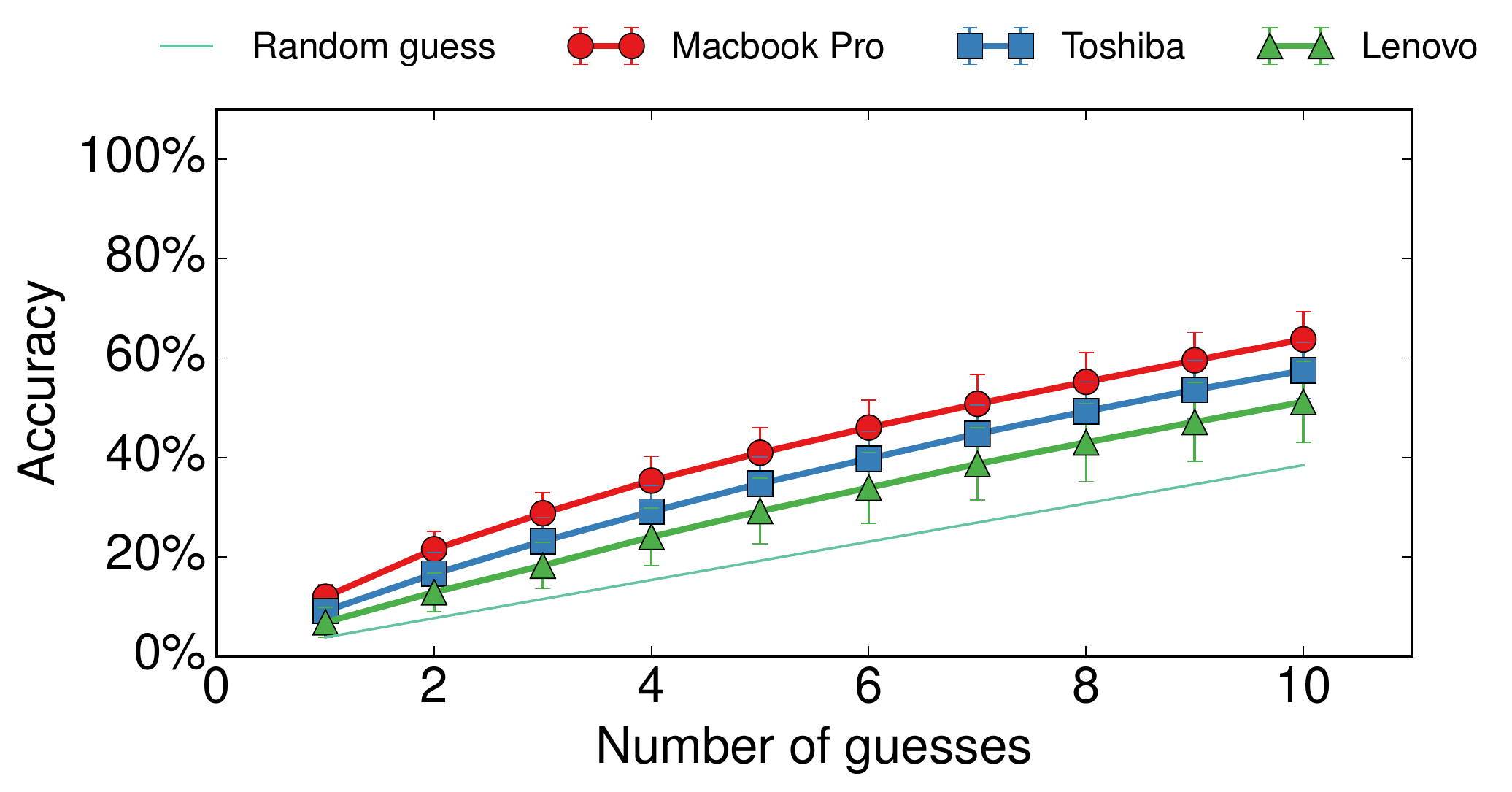}
	\caption[Attack accuracy, \textit{Model Profiling} scenario]{\attack performance -- \textit{Model
			Profiling} scenario, average accuracy.}\label{fig:two_all}
\end{figure}

Results of \attack in this scenario are shown in Figure~\ref{fig:two_all}.  
As expected, accuracy decreases with respect to previous
scenarios. However, especially with Macbook Pro and Toshiba
datasets, we still have an appreciable advantage from a random guess
baseline. In particular, top-1 accuracy goes from a 178\% improvement from the 
baseline random guess on Lenovo
datasets, to a 312\% improvement on Macbook Pro datasets. Top-5 accuracy goes from 
a 152\% on Lenovo to a 213\% on Macbook Pro.

\begin{figure}[h!]
	\centering
	\includegraphics[width=0.95\linewidth]{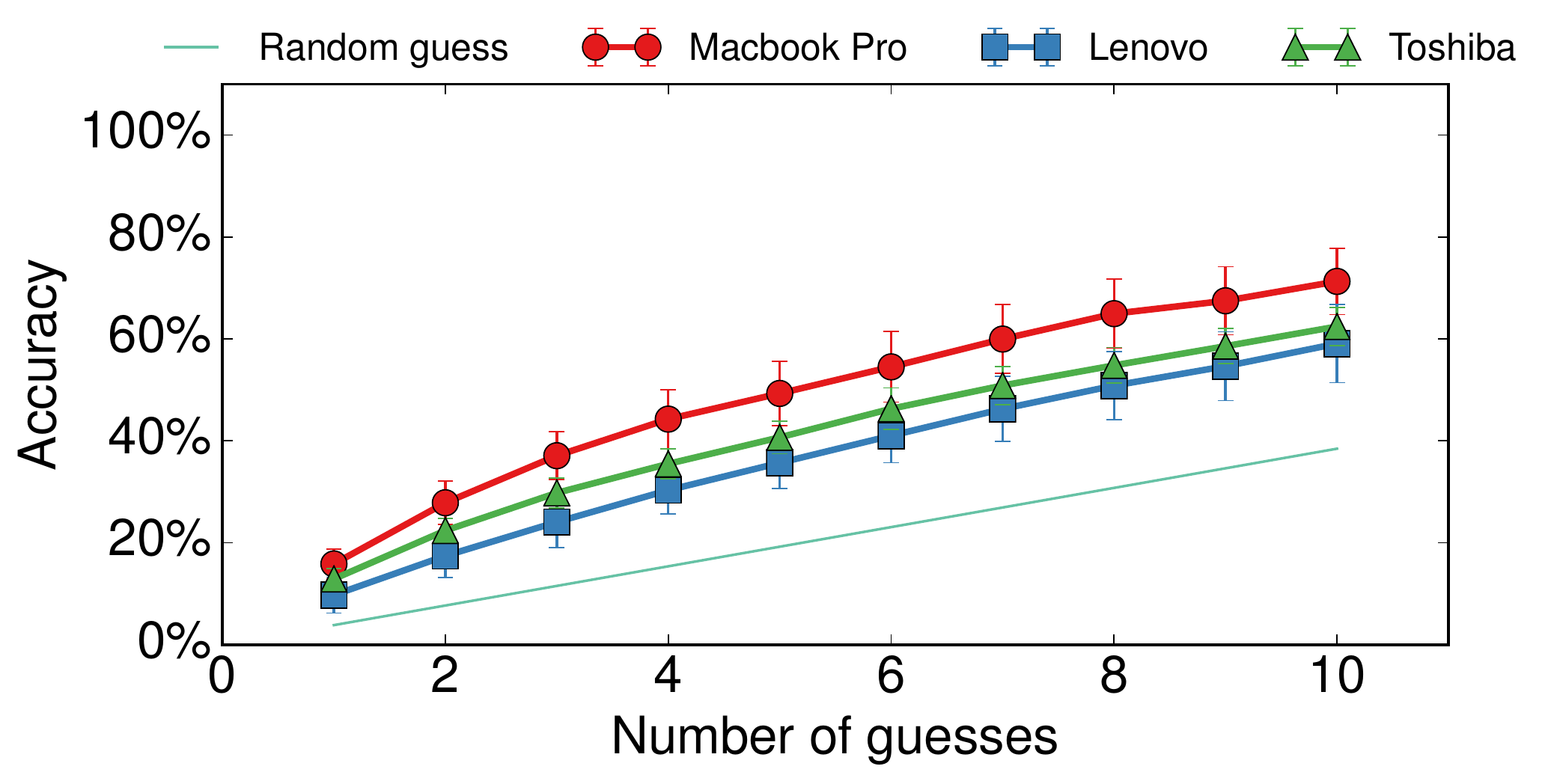}
	\caption[Attack accuracy, \textit{Model Profiling} scenario, ``crowd'' training
	data]{\attack performance -- \textit{Model Profiling} scenario with ``crowd'' training data, average
		accuracy.}\label{fig:crowd}
\end{figure}

To further improve these results, the attacker can use an alternative strategy to
build the training set. Suppose that the attacker recorded multiple users on
a laptop of the same model of the \target\ and then
combines them to form a ``crowd'' training set. We evaluated this scenario as follows: 

We selected the dataset of one user on
a given laptop, as a test set. We then created the training set by combining the
data of other users of the same laptop model. We repeated this
experiment, selecting every combination of user and laptop as a test set, and the
corresponding other users and laptop as a training set. Results reported in
Figure~\ref{fig:crowd} show that overall
accuracy grows by 6-10\%, meaning that this technique further improves 
classifier's detection rate. In particular, 
this increase in accuracy, from 185\% to 412\% (with respect to a
baseline random guess) yields a greater improvement than the approach 
with a single user on the training set.

Results show that \attack\ is still quite viable in a realistic VoIP
scenario, with a target text which is both short and random. Moreover, 
this is possible with little to none specific training data of the victim, i.e., 
the attacker might even have {\em no prior knowledge} of the victim.

\subsection{VoIP-specific Issues}\label{sec:noise}
To conclude the experimental evaluation, we further analyze the impact of 
issues that stem from using VoIP to perform \attack. 
Using VoIP as the attack medium poses additional
challenges to the attacker, such as possible presence of speech on top of
the keystroke sounds. Also, we need to investigate to what extent (if any)
technical features of the SILK codec~\cite{valin2012} degrade performance of \attack. 
For example, this codec reduces audible bandwidth whenever available
Internet bandwidth is low; this operation degrades the sound spectrum. 

We now analyze the impact of variable Internet bandwidth on \attack\
performance, and the impact of voice audio overlaying keyboard emanations,
i.e., the victim talking while pressing keyboard keys.

\subsubsection{Impact of Fluctuating Bandwidth}\label{subsec:skype}
In the experimental setup, both VoIP end-points were connected to a high-speed network. 
However, a realistic call might go over slower or more error-prone network links. 
Therefore, we performed a number of sample Skype calls between the two end-points
while monitoring  network load of the transmitter (i.e., the one producing emanations).

\begin{figure}[htb]
	\centering
	\includegraphics[width=0.95\linewidth]{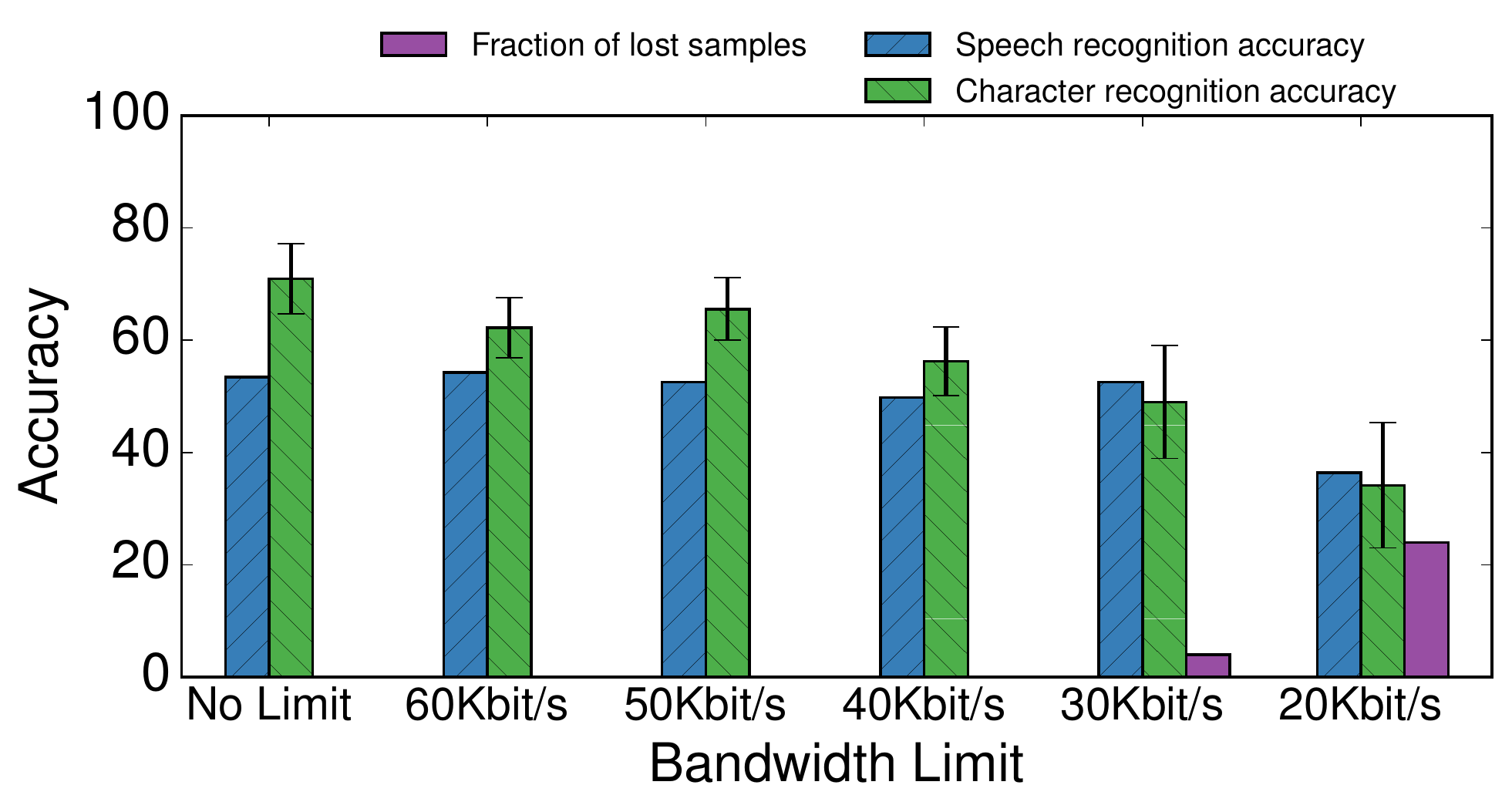}
	\caption[Impact of network bandwidth reduction]{Voice recognition and \attack accuracy, on data acquired through Skype with different connection bandwidths.}
	\label{fig:bandwidth}
\end{figure}

We experimented as follows: we filtered
all data recorded on one Macbook Pro laptop by all the users with
the HP typing style using Skype, together with a five minutes sample of the
\textit{Harvard Sentences}, commonly used to evaluate the quality of VoIP
applications~\cite{rothauser1969}. We initially let the Skype software use the
full bandwidth available, and we measured that the software used an average of
70 Kbit/s without any noticeable packet loss. We subsequently limited the
bandwidth of the transmitting machine at 60 Kbit/s, 50 Kbit/s, 40 Kbit/s, 30
Kbit/s, respectively, 20 Kbit/s. We observed that, with values below 20 Kbit/s,
the quality of the call is compromised, because of frequent
disconnections. \attack with such a small bandwidth is therefore not possible,
and we argue that real users suffering this degradation of service would anyway
not be willing neither able to continue the Skype call. Therefore, we believe
the bandwidths we selected are representative of all the conditions on which we
find the Skype software is able to operate. We then evaluated both the accuracy
of \attack, and the quality of the call by using the voice recognition
software CMU Sphinx v5~\cite{lamere2003} on the Harvard Sentences. We show the
results in Figure~\ref{fig:bandwidth}.

From Figure~\ref{fig:bandwidth}, we can see that, while there is no change to the accuracy of the voice
recognition software until the 20 Kbit/s threshold, the classifier suffers a
noticeable loss at and under 40 Kbit/s. This analysis shows that aggressive
downsampling, and communication errors, can greatly hinder the accuracy of the
attacker on the eavesdropping task, and that a loss of the order of 20\% is to
be expected if the connection speed is very low. We also observe that, at 20
Kbit/s, even if the Skype call is working, many samples of both the speech and
keyboard sounds are lost or irreparably damaged due to the small bandwidth, and
the final quality of the call might be undesirable for the user. However, it is
realistic to assume Skype to be always working at the best possible quality or
almost at the best possible quality, since 70-50Kbit/s are bandwidths that are
small enough to be almost guaranteed.

\subsubsection{The Impact Of Voice}\label{subsec:voice}
In the experiments we described so far, we did not consider that the victim 
can possibly be talking while he types the target text. However, in a VoIP call, 
this can happen frequently, as it is probable that the victim is talking while he
types something on the keyboard of his \target. We evaluated the impact of
this scenario as follows: we considered all the data of one user on the Macbook
Pro laptop, consisting of 260 samples, 10 for every class, in a 10-fold
cross-validation scheme. For every fold, we performed feature selection on the
train data with a Recursive Feature Elimination algorithm, and we then overlapped
the test data with a random part of a recording of some Harvard Sentences with
the pauses stripped out (so that the recording always has some voice in it). To
account for the random overlap, we repeated the process 10 times, to have the
keystroke sound overlap different random phonemes. We then evaluated the mean and
standard deviation of the accuracy of the classifier. 

\begin{figure}[h!]
	\centering
	\includegraphics[width=0.95\linewidth]{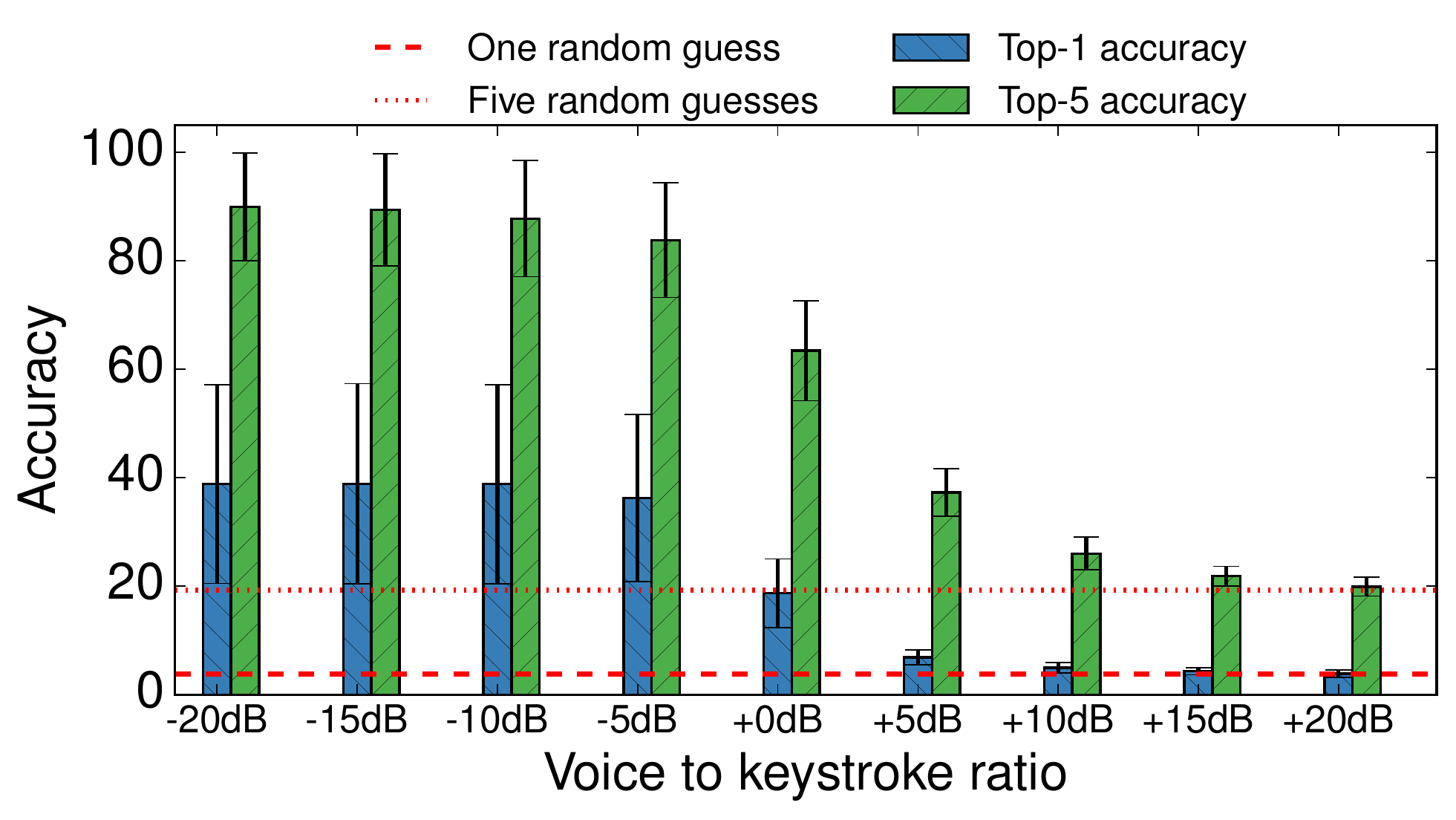}
	\caption[Attack accuracy, voice over keystroke sound]{\attack performance -- average accuracy, overlap
		of keystroke sounds and voice, at different relative intensity.}
	\label{fig:voice}
\end{figure}

We repeated the described experiment with different relative intensities of the voice
against the intensity of the sound of the keystrokes. We started at -$20$dB,
meaning that the keystrokes are $20$dB louder than the voice of the speaker, and
evaluated progressive steps of 5dB, until we had the voice of the speaker $20$dB
louder than the keystrokes. We performed this scheme on the data for all users on
the Macbook Pro laptop, with Touch typing and data filtered with Skype. We show
the results in Figure~\ref{fig:voice}.

We observe that, from $-20$dB until $0$dB, \attack does not suffer almost any
performance loss, and then the accuracy rapidly decreases, until it reaches the
random guess baseline at $+20$dB. We explain both the positive and the negative
results with the phenomenon of auditory masking~\cite{wegel1924}, where only the most
powerful tone among all the tones at a given frequency is audible. In our case,
the greater the difference between the intensity of the sound of the keystroke
and of the voice, the more only the frequencies of the louder sound will be
audible. However, it is realistic to assume that the speaker will talk at a
reasonable volume during the Skype call. Given that the keystrokes are very
loud when recorded from a laptop microphone (sometimes almost peaking the
headroom of the microphone), it is unlikely that the victim will talk more than
5dB louder than a keystroke sound. These results therefore show that the victim
speaking does not prevent the attacker to perform \attack.

\subsection{S\&T Practical Applications}\label{subsec:applications}
We now consider two practical applications of the results of \attack:
understanding words, and cracking random passwords. In particular, if the victim
is typing English words, we analyze how \att\ can help understanding such
words. If the victim is typing a random password, we show how \attack can
greatly reduce the average number of trials required in order to crack it, via
a brute force attack. In the following, we report the results of these practical 
applications on the \textit{Complete Profiling} scenario, and on the 
\textit{Model Profiling} scenario.

\subsubsection{Word Recognition} 
To evaluate how \att\ helps understanding the words that the victim typed,
we proceeded as follows. We drew a number of random words from an English
dictionary; we call such words \textit{actual words}. For each \textit{actual
  word}, we reconstructed its typing sound combining the sound samples of each
letter in the actual word. We used the sound sample of the letters we collected
in Section~\ref{subsec:datasets}. We then performed \attack, to obtain the top-5
predictions for each letter of the actual word, and we created a set of
\textit{guessed words} with the predicted letters. We then calculated the 
error between the \textit{actual word} and the most probable \textit{guessed
  word}, i.e., Hamming distance / length of the word. We tested 1000 random
words for each of the datasets. On the \textit{Complete Profiling} scenario, we
obtain an average error of 9.26\% characters for each word ($\pm$ 8.25\%),
that goes down to 2.65\% ($\pm$ 5.90\%) using a simple spell checker, who is
able to correct most of the errors. We find this trend independent of the word
length. On the \textit{Model Profiling} scenario, we obtain an average error of
60.79\% characters ($\pm$ 9.80\%), down to 57.76\% ($\pm$ 11.50) using spell
checking techniques.  These results are indicative of the possible applications
of \attack, and can be greatly increased with the use of more powerful spell
checking techniques, Natural Language Processing techniques, and crowd-sourced
approaches (e.g., Google Instant).

\subsubsection{Password Recognition}
Secure passwords that prevent dictionary attacks are random combinations of
alphanumeric characters. In order to understand how \attack helps in cracking such
random passwords, we analytically study the speed-up of an improved brute-force scheme
that takes advantage of our results. In particular, the scheme is as follows: given the $x$ guesses of \att\
for each of the $n$ characters of the target password, we first consider all the $x^n$
combinations of such characters. We then assume that the set of $x$ guesses of
the first character was wrong, and subsequently consider all the other
characters. When we finish considering that one set of guesses was wrong, we
consider all the combinations of two wrong guesses (i.e., first and second sets
of guesses were wrong, first and third sets were wrong, up to the seventh and
eighth sets). We repeat this scheme until we finally try the combinations where
the classifier was always wrong. This brute-force scheme leverages the
probability of success of \att\ to minimize, on average, the required time
to crack a password. If we consider a target password of 10 lowercase characters of the
English alphabet, a regular brute-force scheme requires requires
$\frac{(26)^{10}}{2} = 8.39 \cdot 10^{13}$ guesses to have 50\% probability. On
the \textit{Complete Profiling} scenario, that we recall has an average top-5
accuracy 
of more than 90\%, we only need $9.76 \cdot 10^6$ tries to
have 50\% probability. This corresponds to a very high average speedup of
$10^{7}$, and an entropy reduction of more than 50\%. On the \textit{Model Profiling} 
scenario, where we have a top-5 accuracy
around 40\%, we need $7.79 \cdot 10^{12}$ tries to reach
50\% probability of cracking the password, which is still one order of magnitude
better than plain brute-force attacks, on average. There is similar tendency if
the attack guesses ten characters for every character of the password.

\section{Possible Countermeasures}\label{sec:countermeasures}
In this section, we present and discuss some potential countermeasures and analyze their 
efficacy in preventing \att\ and other attacks that use statistical properties
of the sound spectrum.

One simple countermeasure is a short ``ducking'' effect, a technique that drastically
lowers microphone volume and overlaps it with a different sound, whenever a 
keystroke is detected. However, this approach can degrade voice call quality. Ideally, 
an effective countermeasure should be minimally intrusive and affect only keystroke sounds.

A less intrusive countermeasure that might work against all  techniques that use sound spectrum 
information, is to perform short random transformations to the sound whenever a keystroke is detected. 
One intuitive way to do this is to apply a random multi-band equalizer over a number
of small frequency bands of the spectrum. This allows us to modify the
intensity of specific frequency ranges, called ``bands''. Each band should be
selected at random and its intensity should be modified by a small random
amount, thus effectively changing the sound spectrum.  
This approach should allow the speaker's voice to remain intelligible.

\begin{figure}[htb]
	\centering
	\includegraphics[width=0.95\linewidth]{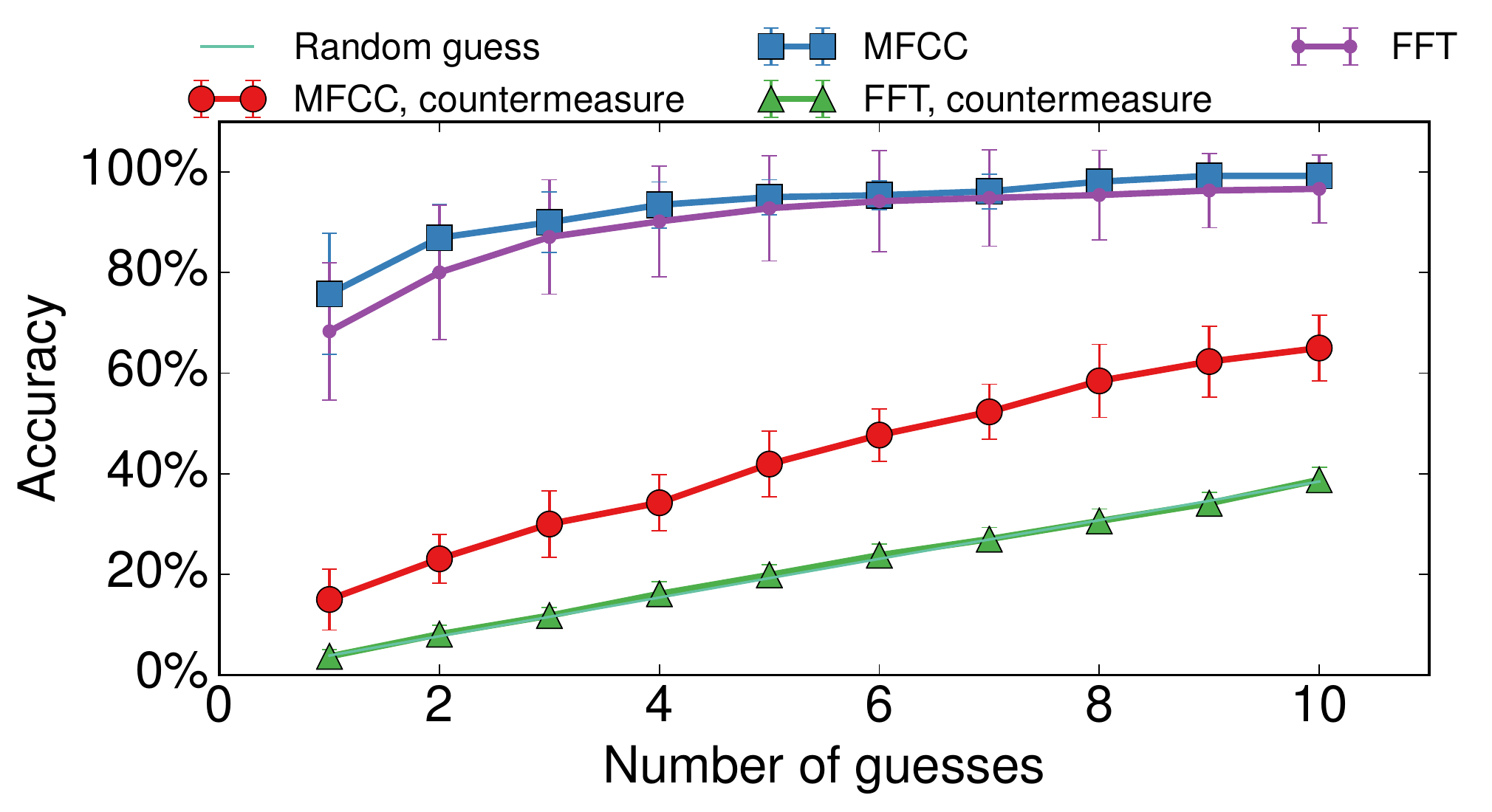}
	\caption{Average accuracy of single key classification against a random 
		equalization countermeasure.}
	\label{fig:countermeasure}
\end{figure}

To show the efficacy of this countermeasure, we ran the following experiment:
we considered all data recorded on the Macbook Pro laptop, one user
at a time, in a 10-fold cross-validation scheme. For every fold, we applied a
multiband equalizer with 100 bands to the test data only, where each band has a
random center between 100 Hz and 3000 Hz, a very high resonance $Q$ of 50, and a
random gain between -5dB and +5dB. We then tried to classify these samples
using both MFCC and FFT features, in order to see if such countermeasure are
effective even against different spectral features. Results 
in Figure~\ref{fig:countermeasure} show 
\att\ accuracy, with and without the countermeasure, for MFCC and FFT features.

The proposed countermeasure successfully disrupts FFT coefficients, such
as those used in~\cite{Asonov2004,Halevi2012,Halevi2014,martinasek2015}, by
reducing the accuracy of \att\ to the baseline random guess. For MFCC
features, although the countermeasure still manages to reduce the accuracy by 50\%, on average, 
the features remain partly robust to this tampering.

A more simplistic approach is to use software or emulated keyboards, 
i.e., those that appear on the screen and are operated by the mouse.
Similarly trivial ideas include: (1) activating a mute button before typing, or 
(2) not to type at all whenever engaged in a VoIP call.

\section{Conclusions}\label{sec:conclusions}
This paper demonstrated a highly accurate VoIP-based remote keyboard acoustic
eavesdropping attack. We first described a number of practical attack scenarios,
using VoIP as a novel means to acquire acoustic information under realistic
assumptions: random target text and very small training sets, in
Section~\ref{sec:sys}. Then, in Section~\ref{sec:attack} we demonstrated an attack with these
assumptions in mind and carefully selected the tools to maximize its accuracy.
In Section~\ref{sec:evaluation}, we thoroughly evaluated \attack\ using Skype in several
scenarios. Finally, we discussed some potential
countermeasures to \att\ and other attacks that leverage spectral 
features of keyboard sounds, in Section~\ref{sec:countermeasures}.

We believe that this work, due to its real-world applicability, advances the 
state-of-the-art in acoustic eavesdropping attacks. \attack\  
was shown to be both feasible and accurate over Skype, in all
considered attack scenarios, with none or minimal profiling of the victim's
typing style and keyboard. In particular, it is accurate in the \textit{Model
  Profiling} scenario, where the attacker profiles a laptop of the same model as 
the victim's laptop, without any additional information about the victim. This
allows the attacker to learn private information, such as sensitive text or 
passwords. We also took into account VoIP-specific issues --
such as the impact of audible bandwidth reduction, 
and effects of human voice mixed with keystroke audio --
and showed that \att\ is robust with respect to both.
Finally, we discussed some countermeasures and concluded that  \att\ is hard to mitigate. 

\section{Future Work}\label{sec:future}
We believe that our choice of laptops and test users is a representative
sample. The number of tested laptops was in line with related work,
and the number of users was greater. (In fact, related work was based on collected 
data of only one user~\cite{Asonov2004,Halevi2012,Halevi2014,martinasek2015}). 
However, it would be useful to run the experiments on more keyboard models 
(such as external keyboards with switches) and with more
users. This would offer a more convincing demonstration that \att\ works regardless of 
underlying equipment and typing styles. Another important direction
is analyzing the impact of different microphones to collect both training and test data. 

As far as the impact of the actual VoIP software, we focused on 
Skype -- currently the most popular VoIP tool~\cite{skypeusers,skypeminutes,skypemobile}.
We consider it to be representative of other VoIP software, since its codecs 
are used in Opus (an IETF standard~\cite{valin2012}) and employed in many VoIP applications, 
such as Google Hangouts and Teamspeak~\cite{opussupport}.
We believe that other VoIP software is probably vulnerable to \attack. 
We also ran some preliminary experiments with Google Hangouts and the results confirm this assertion. 
However, a more thorough assessment of other VoIP software is needed.

We also plan to improve the accuracy of \attack, especially when \tartxt{} is meaningful,
(e.g., English text) by including Natural Language Processing (NLP) techniques or 
crowd-sourcing approaches.
Finaly, we intend to further explore \att\ countermeasures, 
analyze real-time feasibility of random equalization in the presence
of keystroke audio, evaluate its impact on user-perceived call quality,
 and improve its performance.

{\raggedright \printbibliography}
\flushcolsend

\newpage
\appendix
We now analyze the accuracy of \attack\ in the context of the \textit{Complete Profiling} scenario. 

\section{Further Data Comparisons}\label{sec:appendix1}
\begin{figure}[htb]
	\centering
	\includegraphics[width=0.95\linewidth]{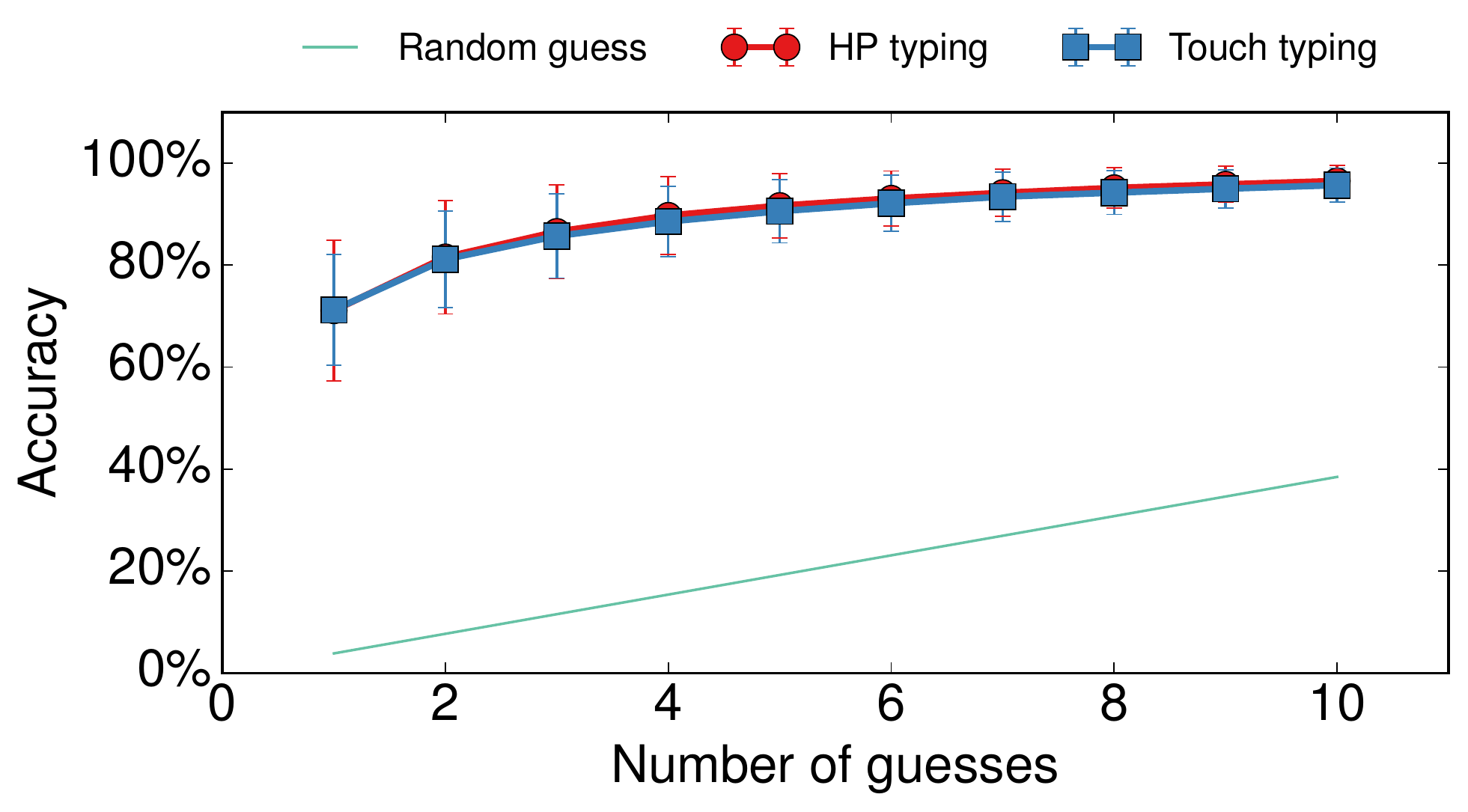}
	\caption{\attack performance -- average accuracy of HP and Touch typing data.}
	\label{fig:cmp_hp_touch}
\end{figure}

\begin{figure}[htb]
	\centering
	\includegraphics[width=0.95\linewidth]{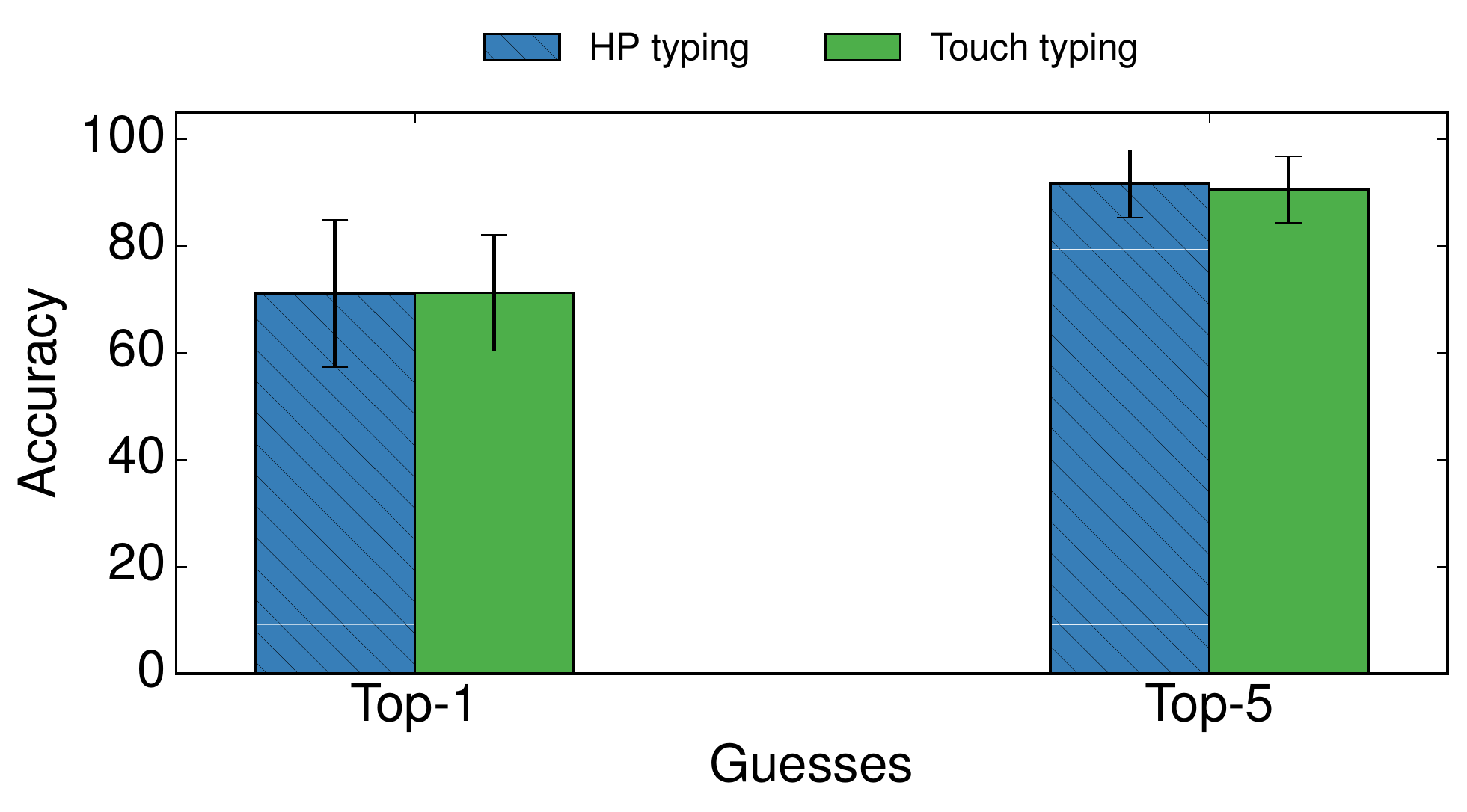}
	\caption{\attack performance -- top-1 and top-5 accuracies of HP and Touch typing data.}
	\label{fig:hp_touch_bars}
\end{figure}

We compare HP and Touch typing data in Figures \ref{fig:cmp_hp_touch} and \ref{fig:hp_touch_bars}. 
Figure \ref{fig:cmp_hp_touch} shows \attack accuracy as a function of the number of guesses, 
and Figure \ref{fig:hp_touch_bars} highlights top-1 and top-5 accuracies. We observe that \attack 
is as accurate with Touch as with HP typing data, within best 4 guesses. From the 5-th guess onwards,
there is a slight advantage with HP typing data; however, the difference is very small -- around 1.1\% 
in the worst case.


\begin{figure}[htb]
	\centering
	\includegraphics[width=0.95\linewidth]{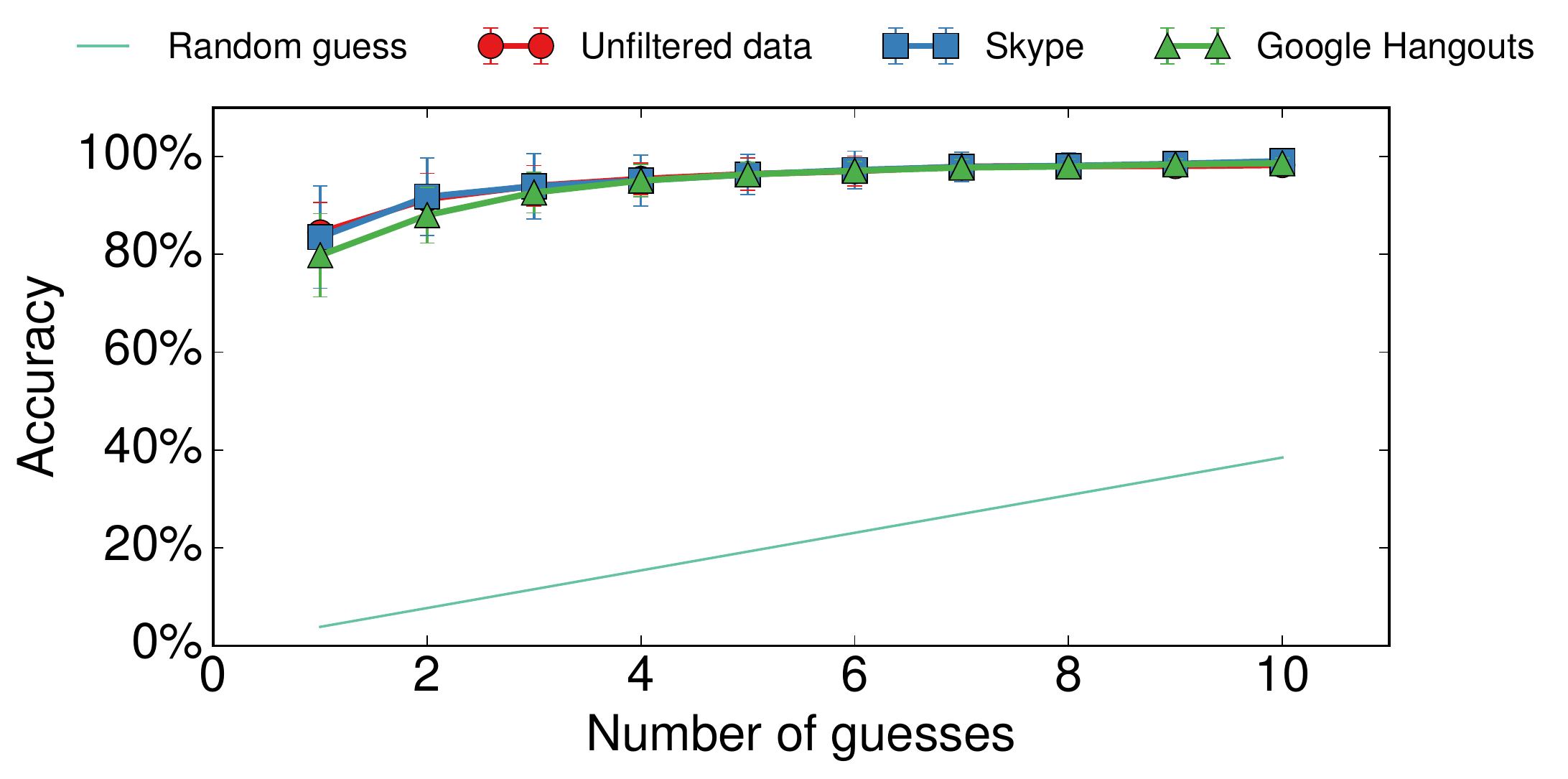}
	\caption{\attack performance -- average accuracy of unfiltered, Skype-filtered and Google Hangouts-filtered data.}
	\label{fig:plain_skype_hangouts}
\end{figure}

\begin{figure}[htb]
	\centering
	\includegraphics[width=0.95\linewidth]{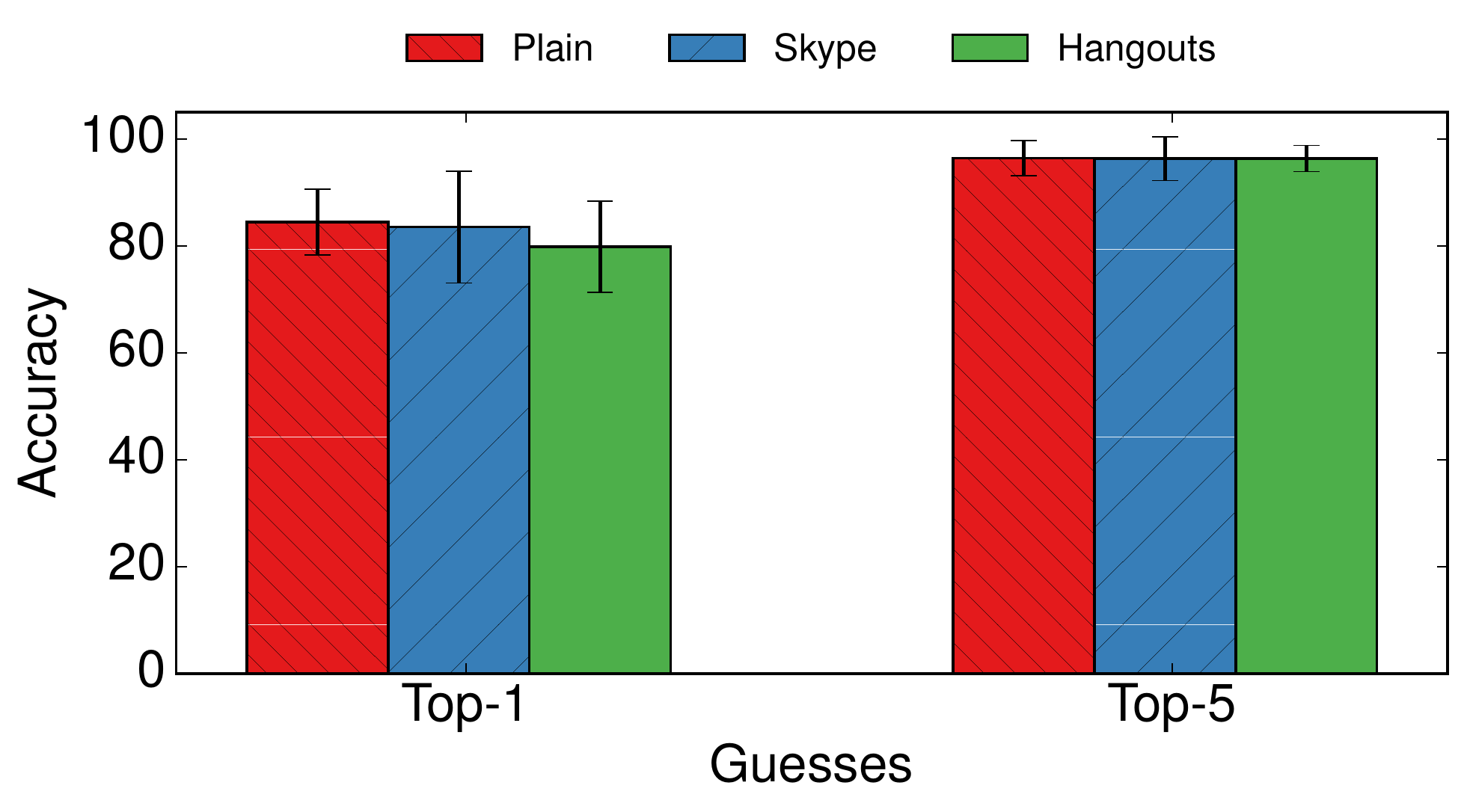}
	\caption{\attack performance -- top-1 and top-5 accuracies of unfiltered, Skype-filtered and Google Hangouts-filtered data.}
	\label{fig:plain_skype_hangouts_bars}
\end{figure}

Next, we compare the following data: unfiltered, Skype-filtered and Google Hangouts-filtered in figures \ref{fig:plain_skype_hangouts} and \ref{fig:plain_skype_hangouts_bars}. Figure \ref{fig:plain_skype_hangouts} 
shows \attack accuracy as a function of the number of guesses, and Figure \ref{fig:plain_skype_hangouts_bars}  
highlights top-1 and top-5 accuracies. Once again, we observe that there is only a small difference in the accuracies 
between unfiltered and Skype-filtered data -- around 1\%. We see a slightly worse top-1 accuracy with Google Hangouts, 
with respect to unfiltered data. This difference of about 5\% gets progressively smaller, and, at top-5, there is no difference
between unfiltered and Google Hangouts-filtered data.

\end{document}